%% file: paper.tex
\documentclass[11pt,a4paper]{article}
\usepackage{jheppub,xcolor}
\usepackage[utf8]{inputenc}
\usepackage{blindtext}
\usepackage{graphicx}
\usepackage{physics}
\usepackage{amsmath}
\usepackage{amsfonts}
\usepackage{amsthm}
\usepackage{amssymb}
\usepackage{caption}
\usepackage{booktabs}
\usepackage{subcaption}
\usepackage{slashed}
\usepackage{bigints}
\usepackage{setspace}
\usepackage{array}
\usepackage{comment}
\definecolor{mygreen}{rgb}{0, 0.5019607843137255, 0}
\definecolor{Red}{rgb}{1.,0.,0.}
\definecolor{Blue}{rgb}{0.,0.,1.}

\usepackage[compat=1.1.0]{tikz-feynman}

\newcommand{\ctGG}{c_{tG}}
\newcommand{\ctpp}{c_{t \varphi}}
\newcommand{\cpGG}{c_{\varphi G}}

\renewcommand{\arraystretch}{1.5}

\title{Impact of SMEFT renormalisation group running on Higgs production at the LHC}

\author[a,b,c]{Fabio Maltoni,}
\author[d]{Giuseppe Ventura,}
\author[d]{Eleni Vryonidou}

\emailAdd{fabio.maltoni@unibo.it} \emailAdd{giuseppe.ventura@manchester.ac.uk} \emailAdd{eleni.vryonidou@manchester.ac.uk}
\affiliation[a]{Dipartimento di Fisica e Astronomia, Universit\`a di Bologna, via Irnerio 46, 40126 Bologna, Italy}
\affiliation[b]{INFN, Sezione di Bologna, Bologna, Italy}
\affiliation[c]{Center for Cosmology, Particle Physics and Phenomenology, Université Catholique de Louvain, Louvain-la-Neuve, Belgium}
\affiliation[d]{Department of Physics and Astronomy, University of Manchester, Oxford Road, Manchester M13~9PL, United Kingdom}

\abstract{
We study the effects of the one-loop renormalisation group running/mixing of the Wilson coefficients in the standard model effective field theory on the predictions for $Hj$, $t\bar tH$ and $HH$ production at the LHC. We focus on a subset of operators closed under QCD corrections and explore the differences between employing a fixed or dynamical scale on the SMEFT predictions for key observables such as the Higgs transverse momentum and the Higgs pair invariant mass. We then study the impact of consistently taking into account renormalisation group effects on the constraints that can be obtained on the Wilson coefficients through current and future measurements at the LHC.

}
\date{}
\begin{document}

\maketitle

\newpage
%%%%%%%%%%%%%%%%%%%%%%%%%%%%%%%%%%%%%%%%%
\section{Introduction}
%%%%%%%%%%%%%%%%%%%%%%%%%%%%%%%%%%%%%%%%%

Absence of evidence of new degrees of freedom in current collider experiments suggests the existence of a substantial mass gap between the electroweak scale and that at which  new physics might reside. In this scenario, the Standard Model Effective Field Theory (SMEFT) \cite{Weinberg:1979sa, Leung:1984ni, Buchmuller:1985jz, 1008.4884} stands out as a robust framework for investigating low-energy signatures of high-scale new physics.  The SMEFT systematically parameterises short-distance deviations from SM predictions through the inclusion of higher-dimensional operators in the Lagrangian while requiring minimal assumptions on the UV nature of beyond the SM physics.

Currently, a comprehensive programme of SMEFT analyses of high-energy experimental data, both inclusive and differential, is in place providing a novel pathway to probe for new physics at collider experiments. This approach could potentially uncover deviations from the Standard Model (SM) or, at a minimum, help pinpoint the scale at which new physics might exist.  The distinctive feature of SMEFT is that it correlates measurements, possibly taking place at different scales. Ongoing efforts aim at SMEFT global interpretations of measurements \cite{2012.02779,Ethier:2021bye, Celada:2024mcf}, i.e., at combining data from  different experiments along with precise theoretical predictions to simultaneously extract the coefficients of large sets of dimension-6 Wilson coefficients. 

In this endeavour, precise theoretical predictions are not only required for the SM but also for the SMEFT. Systematic parallel campaigns for improving both of them are ongoing. In particular, the effort of upgrading SMEFT predictions to higher orders in the QCD and EW couplings is well underway, with QCD corrections automated in Ref.~\cite{2008.11743}. 
In the context of global fits, the interpretation of measurements at different scales and of differential measurements, demands to also properly account for the Renormalisation Group (RG) running and mixing of the SMEFT coefficients. The running and mixing of the Wilson coefficients is determined by the anomalous dimension matrix which is fully known at one loop \cite{1308.2627, 1310.4838, 1312.2014}. The solution of the RG Equations (RGEs) for subsets of dimension-8 SMEFT operators is also available at one-loop \cite{2106.05291, 2108.03669, 2205.03301, 2212.03253, 2307.03187}, and efforts towards extending this to higher loop orders are being pursued by the theory community \cite{1910.05831,2005.12917,2310.19883}.

Several implementations of the one-loop RGEs exist allowing the user to extract the value of the Wilson coefficients at different scales by numerically solving the corresponding system of differential equations \cite{1704.04504, 2010.16341, 1804.05033, 1309.7030, 2210.06838}. Recently, the first implementation of RG running and mixing of operators within a Monte Carlo generator has been presented in \cite{2212.05067},  allowing the computation of SMEFT predictions with either a fixed or dynamical scale on an event-by-event basis. 

A first application of this implementation for high $p_T$ observables  was also discussed in \cite{2212.05067}, where the impact of RG running and mixing on top quark observables, as well as the constraints set on the relevant Wilson coefficients, was computed. Similar investigations were performed in \cite{2109.02987} for the Higgs sector and more recently for the $t\bar t H$ process in \cite{2312.11327} with independent implementations. These first investigations have already demonstrated that potentially significant effects can be observed and motivate the inclusion RG effects in future SMEFT interpretations. 

In this work we aim to further explore the impact of RG running and mixing on a class of Higgs observables. We focus on the transverse momentum spectrum of the Higgs in gluon fusion and in $t \bar t$ associated production, as well as the invariant mass of the Higgs pair in double Higgs production via gluon fusion. These processes play a central role in determining the nature of the top Yukawa coupling, with their combination proving crucial in breaking degeneracies between the relevant operators \cite{1607.05330, 1608.00977,1205.1065}. Additionally, Higgs pair production stands as the main process to explore the Higgs potential at the LHC as reviewed in \cite{1910.00012}. 

We focus on a subset of dimension-6 operators closed under the one-loop QCD RGEs and investigate how different renormalisation scale choices modify the differential distributions. We compare a fixed renormalisation scale, a dynamical scale as well as the case where RG effects are ignored in the predictions. Once the associated theory predictions are obtained with different settings, constraints on the Wilson coefficients can be extracted by comparing to either existing measurements or projections for the future. We select a set of differential observables for which predictions vary most depending on the scale choice and explore how the directions probed by each measurement are modified depending on the scale choice using LHC Run II measurements and projections for the HL-LHC. 

This paper is organised as follows. In Section \ref{sec: setup} we review our calculation setup whilst in Section \ref{sec: diffdist} we explore the impact of RG effects on differential distributions for the Higgs production processes mentioned above. In Section \ref{sec: Fitting} we perform an exploratory fit to determine if constraints set on the Wilson coefficients are significantly affected by the RG effects. We draw our conclusions in Section \ref{sec: conclusions}. 

%%%%%%%%%%%%%%%%%%%%%%%%%%%%%%%%%%%%%%%%%
\section{Theoretical framework}
\label{sec: setup}
%%%%%%%%%%%%%%%%%%%%%%%%%%%%%%%%%%%%%%%%%

With only dimension six operators, the SMEFT Lagrangian takes the form
\begin{equation}
    \mathcal{L}_{\text{SMEFT}}= \mathcal{L}_{\text{SM}}+\sum_i \frac{c_i}{\Lambda^2} \mathcal{O}_i,
\end{equation}
where $\Lambda$ is the scale of new physics and $c_i$ are the dimensionless Wilson coefficients. In the context of this work, we focus on the following three operators:
\begin{equation}
    \begin{split}
        & \mathcal{O}_{t \varphi} = \qty(\varphi^\dagger \varphi- \frac{v^2}{2}) \Bar{Q} \Tilde{\varphi} t + \text{h.c.}, \\ & \mathcal{O}_{\varphi G} = \qty(\varphi^\dagger \varphi- \frac{v^2}{2}) G_{\mu\nu}^a G^{\mu\nu}_a, \\ & \mathcal{O}_{tG} = ig_s (\Bar{Q} \tau^{\mu\nu} T^a \Tilde{\varphi} t ) G_{\mu\nu}^a + \text{h.c.}, 
    \end{split}
    \label{Operators}
\end{equation}
where $\tau^{\mu\nu} = \frac{1}{2} \comm{\gamma^\mu}{\gamma^\nu}$. $CP$-symmetry on the Lagrangian is assumed, such that $\ctpp$ and $\ctGG$ are real, while $\cpGG$ is real as expected by the Hermiticity of the respective operator. The first operator is known as the dimension six Yukawa operator which modifies the Yukawa coupling while also giving rise to higher multiplicity top-Higgs vertices. The second is the Higgs-gluon contact interaction, which, as the name suggests, generates purely SMEFT Higgs-gluon vertices. The last one is the chromomagnetic dipole operator, which modifies the top-gluon interaction and produces gluon-top-Higgs interactions. This set of operators plays a crucial role in the production of the Higgs in the main production channel, gluon fusion, as well as in top associated production \cite{1607.05330, 1612.00283, 1708.00460, 1806.08832}.

Upon renormalisation \cite{1205.1065, 1308.2627, 1310.4838, 1312.2014, 1607.05330, 1612.00283}, these operators run and mix as dictated by the
RGEs, which for the SMEFT take the usual form
\begin{equation}
    \frac{\dd c_i(\mu)}{\dd \log \mu^2} = \gamma_{ij} c_j(\mu),
    \label{eq: RGEs}
\end{equation}
where $\gamma_{ij}$ is the anomalous mass dimension matrix and $\mu$ is the SMEFT renormalisation scale. The matrix $\pmb{\gamma}$ is computed perturbatively in the SM couplings. In this work, we focus on the one obtained through QCD corrections so that the expansion reads
\begin{equation}
    \gamma_{ij}=\sum_k \qty(\frac{\alpha_s}{4 \pi})^k \gamma_{ij}^{(k)}.
\end{equation}
With this assumption, we can extract the part relevant to the considered operators, that has the form
\begin{equation}
    \gamma_{ij}^{(1)} = \frac{1}{3} \begin{pmatrix} -24 & 96y_t & 96y_t^2 \\ 0 & -6\beta_0 & 12y_t \\ 0 & 0 & 4 \end{pmatrix},
    \label{eq: anomalous dimension}
\end{equation}
where $y_t = \dfrac{\sqrt{2} m_t}{v}$ and $\beta_0 = 11-\frac{2}{3}n_f$. The coefficients are ordered as $\{\ctpp, \cpGG, \ctGG \}$.

The solution to the RGEs can be expressed as
\begin{equation}
    c_i(\mu) = \Gamma_{ij} (\mu, \mu_0) c_j(\mu_0),
    \label{eq: Running SMEFT}
\end{equation}
with $\mu_0$ the initial reference scale. The $\pmb{\Gamma}$ matrix can be computed order by order in $\alpha_s$, which at first order takes the form
\begin{equation}
    \pmb{\Gamma}(\mu,\mu_0) = \exp(\int_{\mu_0}^\mu \dd \log \mu' \pmb{\gamma})= \exp(\int_{\mu_0}^\mu \dd \log \mu' \frac{\alpha_s(\mu')}{4 \pi}\pmb{\gamma}^{(1)}),
\end{equation}
and by using the known one-loop beta function for the strong coupling we can derive
\begin{equation}
     \pmb{\Gamma}(\mu,\mu_0) = \exp[\frac{1}{2\beta_0} \log(\frac{\alpha_s(\mu_0)}{\alpha_s(\mu)}) \pmb{\gamma}^{(1)}].
\end{equation}

In practice, different choices for the reference scale $\mu_0$ can be made. For instance, it can represent the scale of matching with a UV theory, from which the system is then evolved down to lower energies, or the natural scale for the experimental observable considered so that running towards higher scales would be needed to match the SMEFT to UV models. In this study, we have selected $\mu_0=1$ TeV as our reference scale. Our focus is on examining how the running of this system, from the initial scale down to a specific value of $\mu$, affects theoretical predictions for key Higgs observables.

\begin{figure}[h]
            \centering
            \includegraphics[width= 0.9\textwidth]{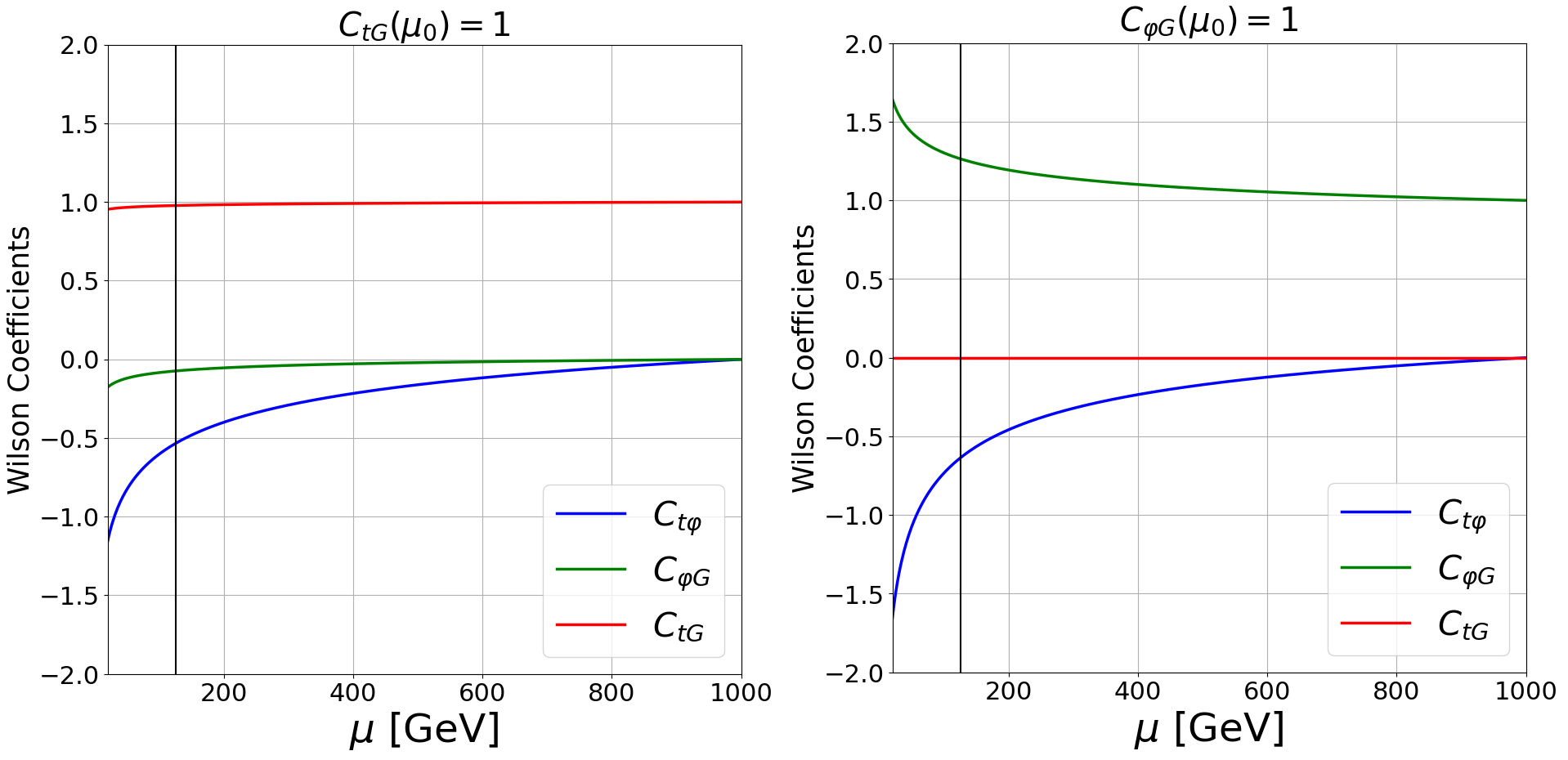}
            \caption{Wilson coefficients as a function of the SMEFT renormalisation scale $\mu$. The reference scale $\mu_0$, as well as $\Lambda$, are set to 1 TeV. On the left, $c_{tG}(\mu_0)=1$ turns on the other two coefficients due to mixing. On the right, $c_{\varphi G}(\mu_0)=1$ only generates $c_{t \varphi}$, while the dipole coefficient is not affected.   The vertical black line indicates the value $\mu=M_H=125$ GeV.
            }
            \label{fig: running plots}
\end{figure}

To set the stage, the dependence of the coefficients on the renormalisation scale $\mu$ is shown in Figure \ref{fig: running plots}, where on the left panel $\ctGG (\mu_0)= 1$, and on the right $\cpGG (\mu_0) = 1$. In the former case, we notice how the mixing causes $\ctpp$ and $\cpGG$ to be activated by the RG due to the dipole coefficient, while in the latter the contact operator $O_{\varphi G}$ only affects the Yukawa operator.\footnote{The case $\ctpp(\mu_0)=1$ is not shown, however $\ctpp$ does not generate the other two coefficients, therefore no mixing occurs. In this case only the running of the coefficient would be seen.}

\subsection{Observables and computational setup}
In the presence of dimension-6 operators, the cross-section is organised as 
\begin{equation}
    \sigma_{\text{SMEFT}} = \sigma_{\text{SM}} + \sum_i \frac{c_i(\mu)}{(\Lambda/1 \ \text{TeV})^2} \sigma_i (\mu) + \sum_{i \leq j} \frac{c_i(\mu) c_j(\mu)}{(\Lambda/1 \ \text{TeV})^4} \sigma_{ij}(\mu),
    \label{eq: EFT cross}
\end{equation} where the second term on the right hand side arises from the interference between diagrams with one operator insertion and the SM and the third one from squaring the EFT diagrams. The dependencies on the factorisation scale $\mu_F$ and the SM renormalisation scale $\mu_R$ are left implicit, yet they are taken into account in the computation\footnote{In principle, the presence of dimension-6 operators alters the running of the SM parameters \cite{1308.2627}, however, this effect is neglected in this work.}. The focus is on the SMEFT renormalisation scale, denoted as $\mu$. As discussed earlier, the scale dependence of the Wilson coefficients is described by the RGEs. However, when considering one-loop diagrams and beyond, a dependence on $\mu$ can also appear in the SMEFT cross-section terms $\sigma_i$ as a consequence of renormalisation. 

Inserting Eq.~\eqref{eq: Running SMEFT} in the expression of the cross-section, we can write the latter in terms of the Wilson coefficients at the reference scale $\mu_0$
\begin{equation}
    \sigma_{\text{SMEFT}} = \sigma_{\text{SM}} + \sum_i c_i(\mu_0) \sigma_i (\mu, \mu_0) + \sum_{i \leq j} c_i(\mu_0) c_j(\mu_0) \sigma_{ij}(\mu, \mu_0),
    \label{eq: EFT cross1}
\end{equation}
where the scale $\Lambda$ has been set to 1 TeV and we define
\begin{equation}
\begin{split}
    & \sigma_i (\mu, \mu_0) = \sum_j \Gamma_{ji}(\mu,\mu_0) \sigma_j(\mu), \\
    & \sigma_{ij} (\mu, \mu_0) = \sum_{kl} \Gamma_{ki}(\mu,\mu_0)\Gamma_{lj}(\mu,\mu_0) \sigma_{kl}(\mu).
\end{split}
\end{equation}
By parameterising  observables (cross-sections and distributions) in this manner, it is possible to extract constraints on the coefficients at the same given scale, but accounting for running and mixing effects in different scenarios.

This paper will focus on Higgs observables in the SMEFT framework, specifically the single Higgs production as a function of its transverse momentum $p_T^H$ and the Higgs pair invariant mass $M_{HH}$ differential cross-section. Several measurements of the Higgs transverse momentum distributions are inclusive, {\it i.e.}, they get contribution from different Higgs production modes. The main one at the LHC, namely the gluon fusion, is sensitive to insertions of the operators in Eq.~\eqref{Operators}. 
At the same time, a contribution from Higgs production in association with a top-antitop pair ($t\bar tH$) is also present and this is also sensitive to the same operators. For the Higgs pair invariant mass distribution we focus on the gluon fusion channel. The  Feynman diagrams with a single SMEFT insertion for the processes we consider are shown in Figure~\ref{fig: FeynmanDiagrams}.

We note here that in addition to the three operators in Eq.~\eqref{Operators}, the processes we consider are affected at the production level by the operator: 
\begin{equation}
O_{d \varphi}= \partial_\mu (\varphi^\dagger \varphi) \partial^\mu (\varphi^\dagger \varphi),
\end{equation} which rescales all Higgs couplings to fermions and gauge bosons. Additionally, double Higgs production via gluon fusion is the main process to probe Higgs trilinear coupling at the LHC \cite{1910.00012} which is modified by $O_{d \varphi}$ and additionally 
\begin{equation}
 O_\varphi = \left( \varphi^\dagger \varphi - \frac{v^2}{2} \right)^3. \label{eq:O6}
\end{equation}
As these two operators do not run under QCD, we will only focus on the operators in Eq. \eqref{Operators} for our inclusive and differential results. We will revisit the operator $O_\varphi$ to determine the impact of RGE effects in the determination of the Higgs self-coupling in Section \ref{sec:fit}.

\input{Fdiagrams}

Our calculations include the leading contributions from each operator in Eq.~\eqref{Operators}. While for $t\bar tH$ this is limited to tree-level diagrams, $Hj$ and $HH$ require one-loop computations as shown in Figure~\ref{fig: FeynmanDiagrams}. In the SM, these processes are loop-induced, but in the SMEFT context the introduction of a contact interaction between the Higgs and gluons allows also for tree-level contributions. At the one-loop level in single Higgs production, inserting the dimension-6 Yukawa operator rescales the $t\bar tH$ coupling, leading to a finite amplitude analoguous to the SM one. This is not the case for an insertion of the chromomagnetic dipole operator. As an example, if we consider the single Higgs production process, the dipole operator would produce an amplitude of the form \cite{1612.00283}
\begin{equation}
    \mathcal{M}(g(p_1) + g(p_2) \rightarrow H) = \ctGG 2 \sqrt{2}\frac{\alpha_s m_t}{\pi} \epsilon_\mu (p_1) \epsilon_\nu (p_2) (p_1^\nu p_2^\mu - p_1 \cdot p_2 g^{\mu \nu})F(\tau),
\end{equation}
with $\epsilon_\mu$ and $\epsilon_\nu$ the polarisation vectors of the incoming gluons and $\tau= 4 m_t^2/m_H^2$. The function $F(\tau)$ is derived by the loop integration and in our conventions takes the form
\begin{equation}
    F(\tau) = \Gamma(1+\varepsilon)\qty(\frac{4 \pi \mu^2}{m_t^2})^\varepsilon \qty(\frac{1}{\varepsilon}+f(\tau)+\mathcal{O}(\varepsilon))=\frac{1}{\varepsilon}+\log \qty(\frac{\mu^2}{m_t^2})+f(\tau)+\mathcal{O}(\varepsilon).
    \label{eq: ctGdivergence}
\end{equation}
The insertion of the dipole operator therefore produces a UV pole in the amplitude \cite{1205.1065, 1607.05330, 1708.00460} that results in a dependence on the renormalisation scale $\mu$ where, on the right hand side of Eq.~\eqref{eq: ctGdivergence}, we employed the \(\overline{\text{MS}}\) scheme. In our notation, the function $f(\tau)$ encompasses all the finite terms that are detailed in \cite{1612.00283}. An analogous pole structure, and thus $\mu$ dependence, is therefore present also in $Hj$ \cite{1806.08832} and $HH$, as studied also analytically in the high-energy limit in \cite{2306.09963}.
This divergence is then absorbed by the renormalisation of the operator $O_{\varphi G}$ \cite{1708.00460}, producing the $\gamma^{(1)}_{23}$ entry of the anomalous mass dimension matrix in Eq.~\eqref{eq: anomalous dimension}. As a result, cross-section terms arising from the insertion of $O_{tG}$ will depend on the renormalisation scale $\mu$. 

The cross-section computation has been performed with \texttt{MadGraph5\_aMC@NLO} \cite{1405.0301} paired with the UFO model \texttt{SMEFT@NLO} \cite{2008.11743}. Each term of Eq.~\eqref{eq: EFT cross} has been computed separately, by appropriately setting the values of the coefficients and the order of the SMEFT expansion. 
Thus, the variation of the Wilson coefficients is accounted for on an event-by-event basis with a custom reweighting code. For tree-level processes, such as $t\bar tH$ production, this method has been tested against the results obtained using the implementation of \cite{2212.05067}, where the running is automatically implemented within \texttt{MadGraph5\_aMC@NLO}, finding full agreeement. 

However, as discussed above, the $\mu$ dependence also appears in the cross-section terms when $O_{tG}$ is inserted.
The model \texttt{SMEFT@NLO} allows the user to select $\mu$ in the parameter card but this does not facilitate an event-by-event $\mu$ dependence computation. By employing suitable cuts and choosing the appropriate $\mu$ value, we account for the scale dependence of the cross-section on a bin-by-bin basis. The binning employed for this task was checked to be  sufficiently fine to obtain precise predictions.

Both dynamical and fixed scale scenarios have been implemented for the SMEFT renormalisation scale. In the former case, the setting for $\mu$ is the Higgs transverse mass  $\mu=M_T^H=\sqrt{(p_T^H)^2+M_H^2}$ when considering the $p_T^H$ spectrum, and the Higgs pair invariant mass $\mu=M_{HH}$ for the $M_{HH}$ spectrum. In the fixed scale scenario, the system is run down from $\mu_0=1$ TeV to $\mu=M_H$. The computation is also performed with no running activated, {\it i.e.}, $\mu=\mu_0=1$ TeV. In all computations, the SM scales $\mu_R$ and $\mu_F$ are dynamical and follow the setting of the SMEFT dynamical scale.

\section{RG effects on differential distributions}
\label{sec: diffdist}
In this section, we show the impact of running and mixing of dimension-6 operators on differential distributions. As previously outlined, our focus is on three processes: $Hj$, $t\bar tH$, and $HH$ production at the LHC, with insertions from operators defined in Eq.~\eqref{Operators}. 
The plots shown in this section compare distributions computed with different choices of the SMEFT renormalisation scale $\mu$, with the sole focus on pure SMEFT terms, which are the ones affected by this scale choice. The ratio between the dynamical scale and the fixed one ($M_H$) is displayed in the insets to quantitatively show the difference.

\begin{figure}[!]
            \centering
            \includegraphics[width= 0.9\textwidth]{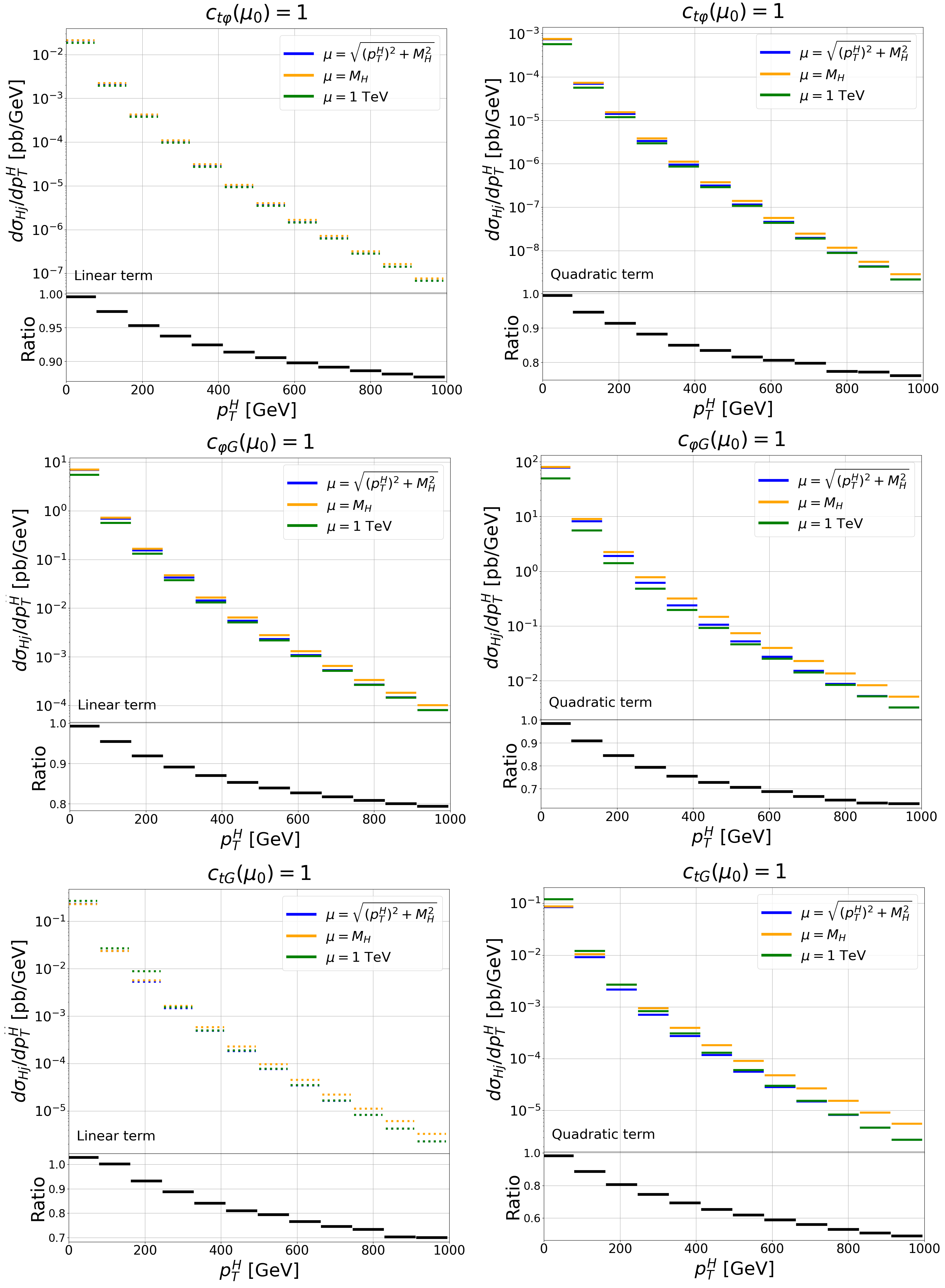}
            \caption{Higgs transverse momentum distribution in $Hj$ production with different settings of the renormalisation scale $\mu$. On the left, only linear terms are considered, on the right only quadratic. The reference scale is $\mu_0=1$ TeV. The two plots in the top panel refer to the case $c_{t \varphi}(\mu_0)=1$, in the middle we have $c_{\varphi G}(\mu_0)=1$, while at the bottom  $c_{t G}(\mu_0)=1$. The other coefficients are set to zero at $\mu_0$. Negative contributions are shown with dashed lines. The insets show the ratio of dynamical over fixed scale ($\mu=M_H$). }
            \label{fig: Hjdistribution}
\end{figure}

\begin{figure}[!]
            \centering
            \includegraphics[width= 0.9\textwidth]{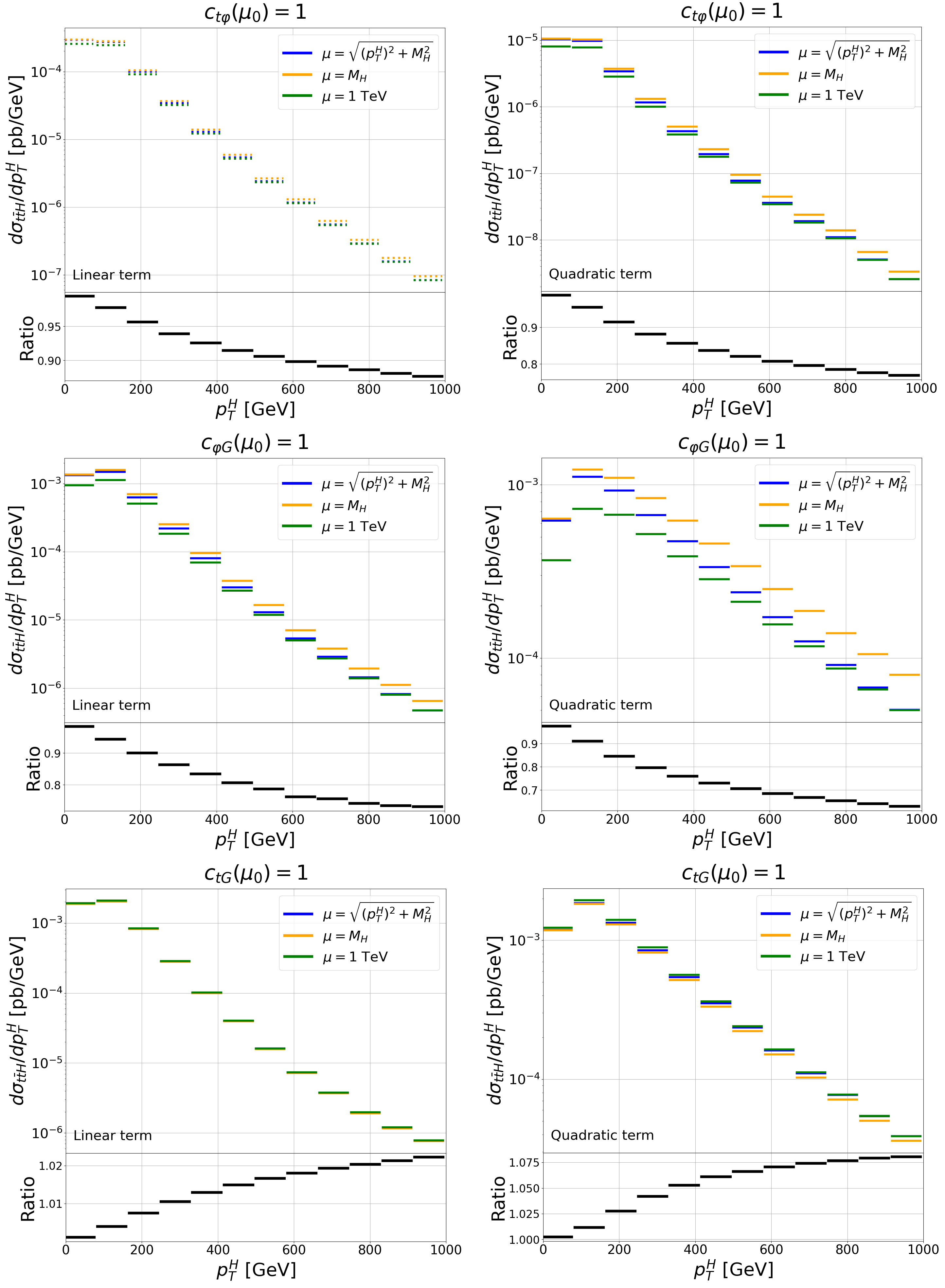}
            \caption{Differential SMEFT cross section for $t\bar tH$ production as a function of $p_T^H$ with different settings of the renormalisation scale $\mu$. On the left, only linear terms are considered, on the right only quadratic. The reference scale is $\mu_0=1$ TeV. The two plots in the top panel refer to the case $c_{t \varphi}(\mu_0)=1$, in the middle we have $c_{\varphi G}(\mu_0)=1$, while at the bottom  $c_{t G}(\mu_0)=1$. The other coefficients are set to zero at $\mu_0$.}
            \label{fig: ttHdistribution}
\end{figure}

\begin{figure}[!]
            \centering
            \includegraphics[width= 0.9\textwidth]{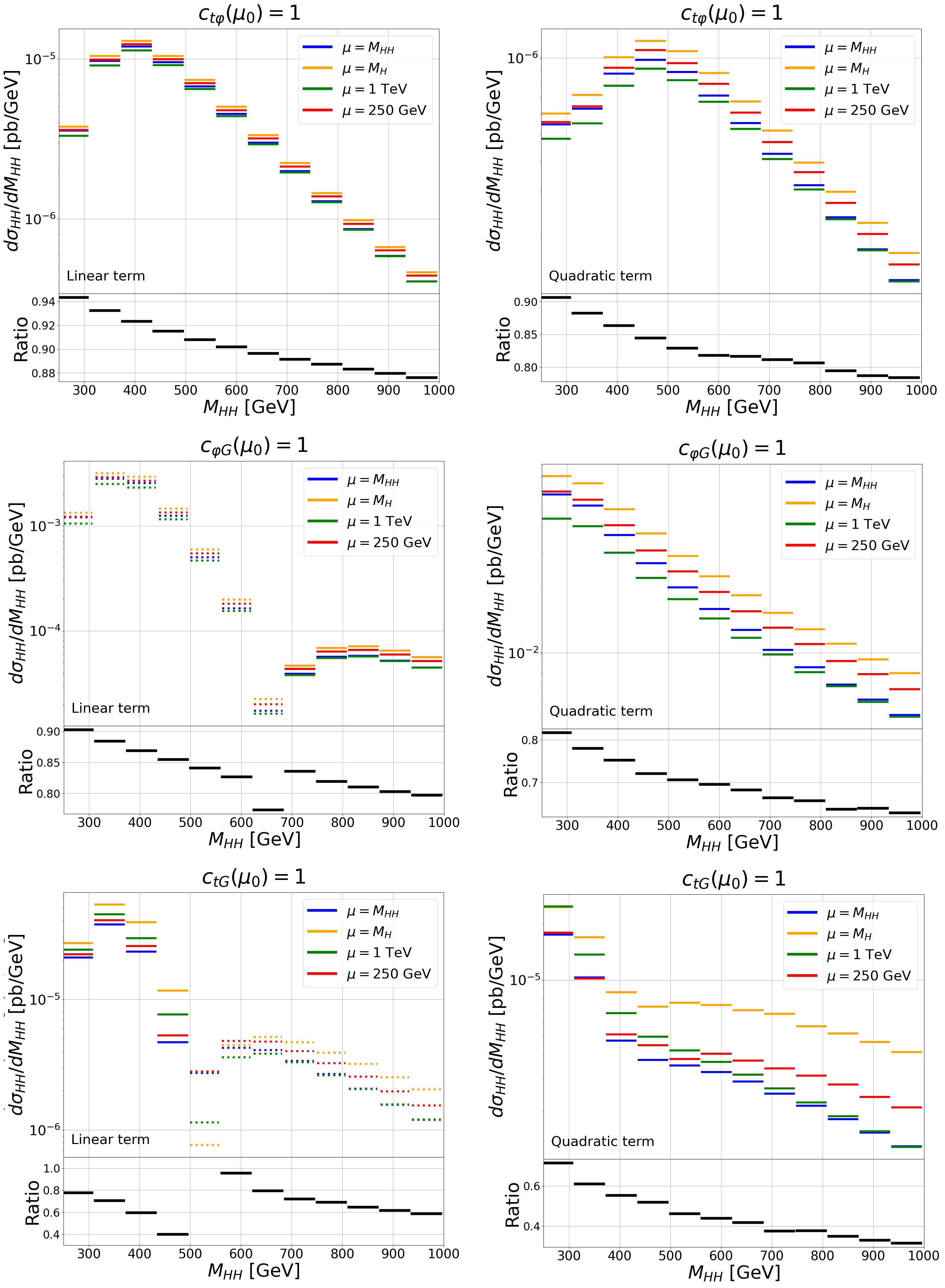}
            \caption{Differential SMEFT cross section for $HH$ production as a function of the Higgs pair invariant mass $M_{HH}$ with different settings of the renormalisation scale $\mu$. On the left, only linear terms are considered, on the right only  quadratic. The reference scale is $\mu_0=1$ TeV. The two plots in the top panel refer to the case $c_{t \varphi}(\mu_0)=1$, in the middle we have $c_{\varphi G}(\mu_0)=1$, while at the bottom  $c_{t G}(\mu_0)=1$. The other coefficients are set to zero at $\mu_0$.}
            \label{fig: HHdistribution}
\end{figure}

The differential plots presented in this section are obtained from setting one of the Wilson coefficients to 1 at the reference scale $\mu_0=1$ TeV. The distributions for $Hj$, $t\bar tH$ and $HH$, are respectively shown in Figures \ref{fig: Hjdistribution}, \ref{fig: ttHdistribution} and \ref{fig: HHdistribution}. The distribution for $\mu=M_H$ indicates the fixed scale scenario, where the system is run down from $\mu_0$ to $M_H$. The one with $\mu = \sqrt{M_H^2+(p_T^H)^2}$ for $Hj$, and equivalently $\mu=M_{HH}$ for $HH$ production, represents the dynamical scale scenario.
%, where the running is performed to the scale of the event for the Wilson coefficients, and to the centre of the bin for the $\mu$-dependent cross-section terms. 
Finally, the case where no running is activated, and the system is kept at $\mu_0$, is indicated with $\mu=1$ TeV. 

In the case of the Higgs transverse momentum spectrum, the dynamical scale spans a range from the Higgs mass value at low $p_T^H$, to approximately 1 TeV at high $p_T^H$. This causes the ratio between the dynamical and fixed scale scenarios to approach one at low transverse momentum as the dynamical scale approaches $M_H$ at low $p_T^H$.
In the case of double Higgs production, the dynamical scale follows the invariant mass spectrum, while the fixed scale set at $\mu=M_H$ is below the threshold of the process. Therefore the ratio plots are not necessarily approaching one as fast as for the $Hj$ process.  For comparison, the computation has been performed with an additional fixed scale scenario, specifically $\mu=250$ GeV, which aligns very well with the dynamical scale close to the production threshold.

Focusing on the plots for $Hj$ and $HH$ production in Figs. \ref{fig: Hjdistribution} and \ref{fig: HHdistribution}, and specifically to the first and second rows, where the dipole coefficient $c_{tG}$ is always zero, we see an effect from running up to $\mathcal{O}(15-40\%)$. First, $c_{t \varphi}$ does not generate the other two coefficients, meaning that the only impact on the distribution is through its own running. In contrast, $c_{\varphi G}$ generates the Yukawa coefficient $c_{t \varphi}$. However, due to the hierarchy between the terms, the resulting mixing is small. In fact, the $O_{\varphi G}$ inserted terms are numerically dominant over the Yukawa inserted ones due to tree-level diagrams. Therefore, also in this case, the impact of scale dependence is mainly caused by the running of $c_{\varphi G}$. In both sets of plots, we also observe that the impact of running is as expected more pronounced for the case of the quadratic contributions, which scale with the square of the coefficient. Hence, the maximum difference between the fixed and dynamical scale is approximately twice as large for the quadratic contribution compared to the linear term for the upper and middle plots. 

When $c_{tG}$ is non-zero, as shown in the lowest set of plots, the effects of scale dependence can have larger impact, up to $\mathcal{O}(50\%)$. The first reason is that, as we previously discussed, the insertion of the dipole operator causes a scale-dependence for the cross-section terms. In addition, $c_{tG}$ will induce the other two coefficients, generating the respective terms in the cross-section. The consequence is a significant mixing of the dipole terms into $O_{\varphi G}$, which induces tree-level contributions to the cross-section. The two combined effects are responsible for the impact we see in the differential distributions. 

For $t\bar tH$ shown in Fig.~\ref{fig: ttHdistribution}, we can also see a variation up to $\mathcal{O}(30\%)$. In this case the mixing effects are always subleading, and the variation is mainly due to the running of the non-zero coefficient at the reference scale. As a consequence, the magnitude of the RG effects is strictly related to the variation of the coefficient. For instance, this explains why the running effects are similar for all the processes when $c_{t \varphi}(\mu_0)=1$ or $c_{\varphi G}(\mu_0)=1$, i.e. when no significant mixing is detected. For the $ttH$ process, when $c_{tG}$ is non-zero the effect of running remains below the 10 percent level. 

\begin{figure}[h]
            \centering
            \includegraphics[width= 0.9\textwidth]{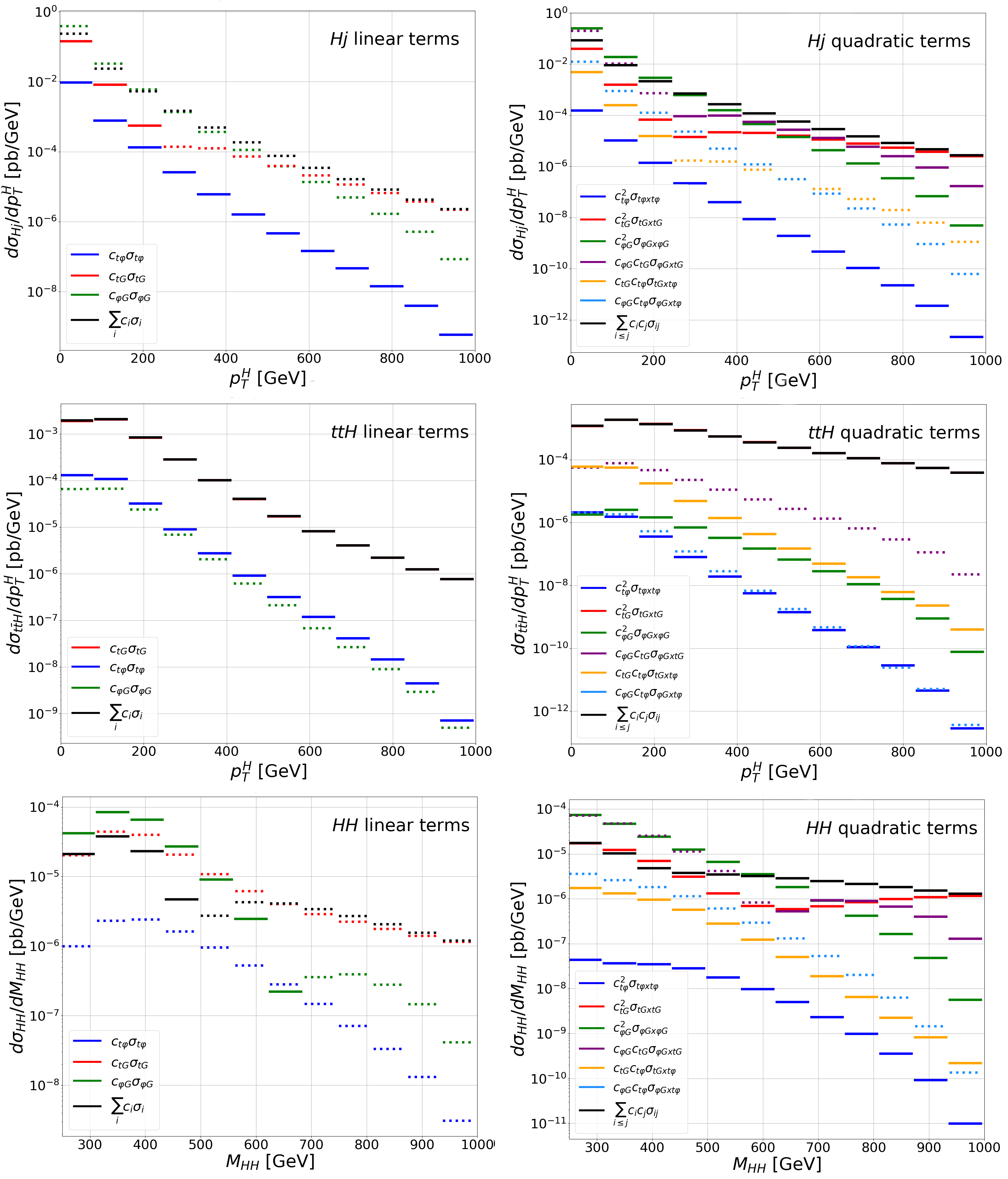}
            \caption{Separate contributions to the differential SMEFT cross section for $Hj$ (top panel), $t\bar tH$ (middle), $HH$ (bottom panel) in the dynamical scale scenario. On the left, linear terms are considered, on the right only quadratic. $c_{tG}(\mu_0)=1$ while the other two coefficients, set initially to zero, are turned on by the running.}
            \label{fig: Splitting Plots}
\end{figure}

Quantitatively, we can show the mixing effects in the dynamical scale scenario by decomposing the cross-section into different contributions for $c_{tG}(\mu_0)=1$. These are shown in Figure \ref{fig: Splitting Plots}. In the case of $HH$ and $Hj$ we notice that whilst in the tails the total cross-section is dominated by the $O_{tG}$-inserted term, the $O_{\varphi G}$ ones generated by the RGEs are dominant at low energies. The decomposition is more complex for the quadratic contributions as once we depart from the original scale six contributions are generated. The final distribution is then the result of the relative signs and sizes of these six terms. In contrast to the loop induced processes, for $t\bar tH$ and the case of $c_{tG}(\mu_0)=1$, as shown in the middle panel of Figure \ref{fig: Splitting Plots} the contributions induced by mixing are negligible, with $\sigma_{tG}$ in red overlapping with the sum of all contributions in black. Moreover, the variation of $c_{t G}$ in the considered scale range is of the order of 1\% as shown in Figure \ref{fig: running plots}, and a consistent variation was also found for the differential cross-section, as depicted in Figure \ref{fig: ttHdistribution}.

It should be noted that the results shown for the differential distributions aim to demonstrate the impact of RG running for different scale choices. The final goal is to determine the impact of these effects on the extraction of bounds for the Wilson coefficients, which will be discussed in the next section. 

\section{Constraints on Wilson coefficients}
\label{sec:fit}
In this section, we aim to assess the impact of RG running and mixing on the bounds of Wilson coefficients by performing a set of toy fits. To demonstrate the potential impact of these effects, we select a subset of differential Higgs observables, whose predictions vary most depending on the scale choice, the $p_T^H$ and $M_{HH}$ distributions. In our theoretical framework, the former receives SMEFT contributions from the gluon fusion $Hj$ and $t\bar tH$ production modes, while the latter is determined by double Higgs production via gluon fusion. For the Higgs transverse momentum experimental measurements already exist from Run II at the LHC. We will employ those to constrain our operators with present data and combine projections for the Higgs transverse momentum and the Higgs pair invariant mass to extract constraints based on the full HL-LHC dataset. 

Whilst the gluon fusion processes are the main contributors to the respective spectra at the LHC, additional production modes should be accounted for if the experimental setup is sufficiently inclusive. This is especially true for the tails of the $p_T^H$ spectrum, which receive significant contributions from the vector boson fusion VBF and Higgs-Strahlung $VH$ production modes \cite{2005.07762}. However, up to NLO, the electroweak Higgs production processes are not sensitive to the operators in Eq.~\eqref{Operators}, therefore we will consider them to behave as their respective SM predictions. Regarding the Higgs pair invariant mass spectrum that we consider at the HL-LHC, the other production modes are subleading with respect to the gluon fusion \cite{1910.00012}, and while, for example, the double Higgs production in association with a top-antitop pair is sensitive to the SMEFT operators considered in this work, the related cross-section is negligible. We also expect that with sufficiently exclusive experimental selections one can isolate the contribution originating from gluon fusion.  

Throughout this section, constraints are extracted on the value of the  Wilson coefficients at the reference scale $\mu_0=1$ TeV for different SMEFT renormalisation scale scenarios, as also discussed in Section \ref{sec: setup}. 
\paragraph{Fit input} We start by fitting the Higgs transverse momentum spectrum to the current available data presented by ATLAS~\cite{ATLAS:2019mju, ATLAS:2021nsx} and CMS~\cite{2006.13251, 2103.04956} for the differential Higgs cross section in different $p_T^H$ regions. In our analysis, we are interested mostly in the boosted regime, which we expect to be most sensitive to the different shapes originating from the different operators~\footnote{In a global analysis an additional constraint on the Wilson coefficients will arise from the total Higgs production cross-section. At the inclusive level, it is well known \cite{1205.1065,1608.00977} that the inclusive cross-section constrains a linear combination of the $c_{t\phi}$ and $c_{\phi G}$ coefficients}. We thus focus on $p_T^H > 200$ GeV which is expected to provide most of the sensitivity to the SMEFT effects. At high $p_T^H$ the data are obtained through the Higgs decaying into a $b \Bar{b}$ pair, while at low $p_T^H$ the differential cross-section is reconstructed through $H\rightarrow\gamma \gamma$~\footnote{We note here that we do not consider the impact of the operators on the partial and total Higgs decay widths. } and $H\rightarrow Z Z^*\rightarrow 4 \ell $ decay channels. 

Specifically, the data provided by ATLAS is inclusive of all production modes, while the data from CMS only refers to the gluon fusion mode, where the others are subtracted under the assumption of being SM-like. The signal strength values in the differential regions relevant to our analysis are shown in Table \ref{tab: currentdata}. 
\begin{table}[h]
\small
    \begin{subtable}{\linewidth}
    \centering
        \begin{tabular}{|c|c|c|c|c|c|}
          \hline $p_T^H$ & $[200,350]$ & $>350$ & $[450,650]$ & $[650,1000]$ \\  \hline 
          $\sigma/\sigma_{\textrm{SM}}$&  $1.15 \pm 0.2$  &$1.0 \pm 0.4$ & $-2.9 \pm 4.7$ & $4.8 \pm 6.4$ \\ \hline 
    \end{tabular}
    \caption{}
    \vspace{0.3cm}
    \end{subtable}
    \begin{subtable}{\linewidth}
    \centering
    \begin{tabular}{|c|c|c|c|c|c|}
        \hline $p_T^H$ &  $>200$ & $[300,450]$ & $[450,650]$ & $[650,1000]$ \\  \hline 
      $\sigma/\sigma_{\textrm{SM}}$&     $0.5 \pm 0.5$  &$6.9 \pm 9.1$ & $0.4 \pm 3.2$ & $15.2 \pm 6.0$ \\ \hline 
    \end{tabular}
    \caption{}
    \end{subtable}
    \caption{(a) ATLAS inclusive \cite{ATLAS:2019mju, ATLAS:2021nsx}  and (b) CMS $ggF$ \cite{2103.04956, 2006.13251}  $\sigma/\sigma_{SM}$ values for each $p_T^H$ region with related uncertainties. The top row represents the data binning, in units of GeV.}
    \label{tab: currentdata}
\end{table}

We subsequently perform a fit using projections for the HL-LHC. Both ATLAS \cite{ATLAS:2018ibe} and CMS \cite{CMS:2018qgz} collaborations provided the projected uncertainties for the inclusive $p_T^H$ spectrum. Moreover, we consider separately the differential cross-section for $t\bar tH$ production, with uncertainties taken from the CMS analysis in \cite{CMS:2018rig}. Such uncertainties, as employed in our fit, are shown in Table \ref{tab: HLProjdata}. 

\begin{table}[h]
\small
    \begin{subtable}{\linewidth}
    \centering
        \begin{tabular}{|c|c|c|c|c|c|c|}
          \hline  $p_T^H$ & $[0, 45]$ & $[45, 80]$ & $[80, 120]$ & $[120, 200]$ & $[200, 350]$ & $ >350$ \\  \hline 
          $\Delta \mu$ & $37 \%$  & $28 \%$ & $23 \%$ & $32 \%$ & $30 \%$ & $34 \%$\\ \hline 
    \end{tabular}
    \caption{}
    \vspace{0.3cm}
    \end{subtable}
    \begin{subtable}{0.5\linewidth}
    \centering
    \begin{tabular}{|c|c|c|}
          \hline $p_T^H$ & $[200, 350]$ & $[350, 1000]$ \\  \hline 
          $\Delta \mu$ & $4.5 \%$  & $8.2 \%$  \\ \hline 
    \end{tabular}
    \caption{}
    \vspace{0.3cm}
    \end{subtable}
%    \hspace{-3.45cm}
    \begin{subtable}{0.5\linewidth}
    \centering
    \begin{tabular}{|c|c|c|c|}
          \hline $p_T^H$ & $[200, 350]$ & $[350, 600]$ & $ >600$  \\  \hline 
         $\Delta \mu$ &   $4.4 \%$  &$8.0 \%$ & $24 \%$  \\ \hline 
    \end{tabular}
    \caption{}
    \vspace{0.3cm}
    \end{subtable}
    \begin{subtable}{\linewidth}
    \centering
    \begin{tabular}{|c|c|c|c|}
          \hline $M_{HH}$ & $[250, 500]$ & $[500, 750]$ & $ [750, 1000]$  \\  \hline 
         $\Delta \mu$ &   $36 \%$  &$58 \%$ & $158 \%$  \\ \hline 
    \end{tabular}
    \caption{}
    \end{subtable}
    \caption{Relative uncertainties for projected HL-LHC (a) $t\bar tH$ differential distribution in different $p_T^H$ regions \cite{CMS:2018rig}, (b) inclusive $p_T^H$ spectrum from ATLAS~\cite{ATLAS:2018ibe} and (c) CMS~\cite{CMS:2018qgz}, and (d) Higgs pair invariant mass spectrum \cite{1902.00134}. Top rows represent the data binning in units of GeV.}
    \label{tab: HLProjdata}
\end{table}

For double Higgs production, there are currently no data for the differential distribution, with experimental searches from Run II having set limits on the total cross-section for Higgs pair production of a few times the SM prediction \cite{Portales:2022ggy, 2202.09617, 2209.10910, 2112.11876, 2011.12373, 2206.10657, 2202.07288}. Therefore, we will only perform a toy fit to the SM-spectrum by doing a projection for the HL-LHC. In our implementation, the invariant mass spectrum of the Higgs pair is divided into 3 bins of 250 GeV width, over the interval [250,~1000] GeV. The HL-LHC projection for the uncertainty is only available for the total cross-section \cite{1902.00134}. However, since this uncertainty is dominated by statistics, we can derive the corresponding values for each bin based on the relative SM cross-section values. Ignoring the systematics, the projected  uncertainty on the signal strength is $\Delta \mu_{tot} = 0.31$, implying $\Delta \mu_b = [0.36, 0.58, 1.58]$ in the three considered bins. 
\label{sec: Fitting}

\subsection{Fit methodology}
The fit is performed through a simple $\chi^2$ minimization procedure, allowing for the determination of bounds for the Wilson coefficients. Individual and marginalised fits are performed, where we extract constraints for each of the coefficients, as well as two-dimensional fits, where the bounds are shown on a 2D plane. Then, we can compare the results among different scale choices.

The experimental data for the $p_T^H$ spectrum requires multiple Higgs production modes to be considered in the fit. Let $\sigma$ be the total SMEFT cross-section in a given bin, we can write
\begin{equation}
    \sigma= \sum_a \sigma^a,
\end{equation}
where $a$ runs over the production modes. We take the ratio to the total SM prediction, which reads
\begin{equation}
   R\equiv \frac{\sigma}{\sigma^{SM}} = \sum_a \frac{\sigma^a}{\sigma^{SM}}= \sum_a f_a R_a, 
\end{equation}
where we can define
\begin{equation}
    f_a = \frac{\sigma^{a, SM}}{\sigma^{SM}}, \ \ R_a = \frac{\sigma^a}{\sigma^{a, SM}}.
\end{equation}
The quantity $f_a$ is the weight factor and indicates the contribution of a given production mode to the total spectrum. Regarding the $p_T^H$ spectrum, these factors can be extracted from the experimental analyses. Specifically, ATLAS detected about 50\% of the event being from the gluon fusion mode, a 15\% contribution for $t\bar tH$, while the remaining 35\% from $VH$ and $VBF$ production modes~\cite{ATLAS:2021nsx}. In contrast, CMS reports only a 5\% contribution for $t\bar tH$. Since the CMS dataset subtracted the latter under the assumption of SM behaviour, we will include the SMEFT prediction of $t\bar tH$ only for the ATLAS dataset. The weight factors employed for the HL projections of the inclusive $p_T^H$ spectrum are instead taken from the theoretical prediction of a boosted Higgs production \cite{2005.07762}.

Therefore for the inclusive Higgs transverse momentum spectrum, we fit
\begin{equation}
    \sum_a f_a R_a(c_i(\mu_0)) = \mu_{exp} \pm \Delta \mu_{exp},
\end{equation}
where $\mu_{exp}$ is the data signal strength and $\Delta \mu_{exp}$ the related uncertainty. The $M_{HH}$ spectrum constitutes a simplified case as only one production mode, i.e. the gluon fusion, is taken into account. 

As we employ the signal strengths from the experimental measurements, we use our LO predictions to compute the corresponding SMEFT signal strength. Given that the experimental signal strengths employ the best available SM predictions, we essentially assume that any higher-order corrections affect the SM and SMEFT in the same way. As this is a toy fit aiming to assess the impact of RG effects and not to extract precise constraints on the Wilson coefficients, we consider this to be a reasonable assumption and defer a more realistic fit set-up to future work. 

\subsection{Results}
\begin{figure}[h]
\begin{subfigure}{\linewidth}
    \centering
    \includegraphics[width= \textwidth]{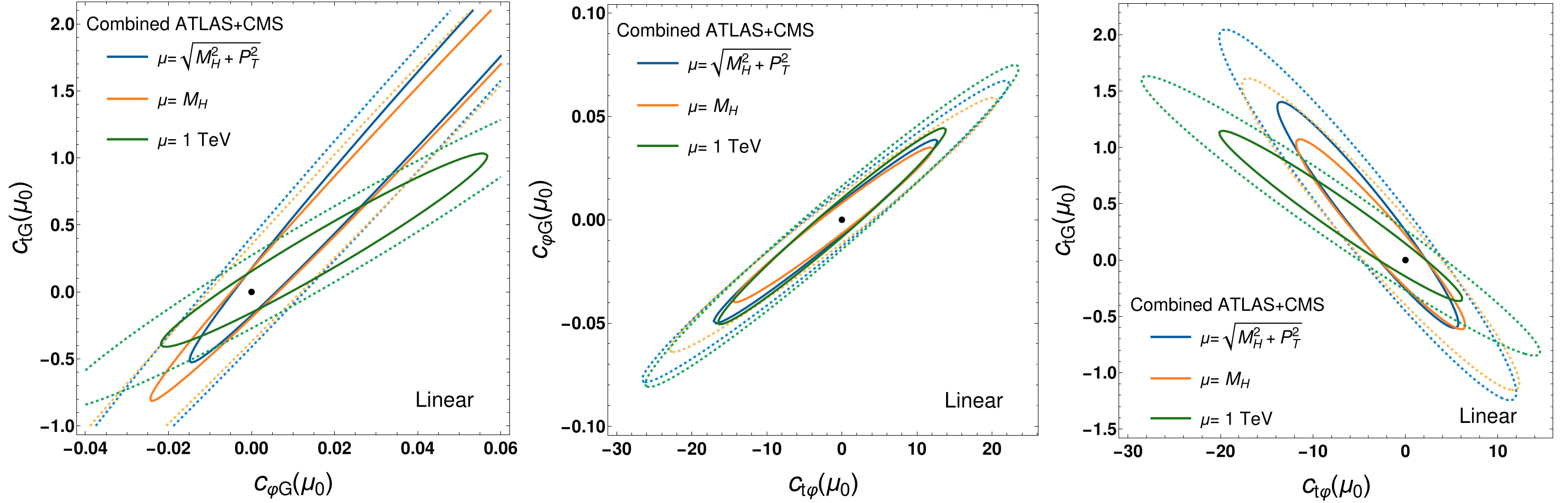}
    \caption{}
    \label{fig: linearFIT}
    \vspace{0.2cm}
\end{subfigure}
\begin{subfigure}{\linewidth}
    \centering
    \includegraphics[width= \textwidth]{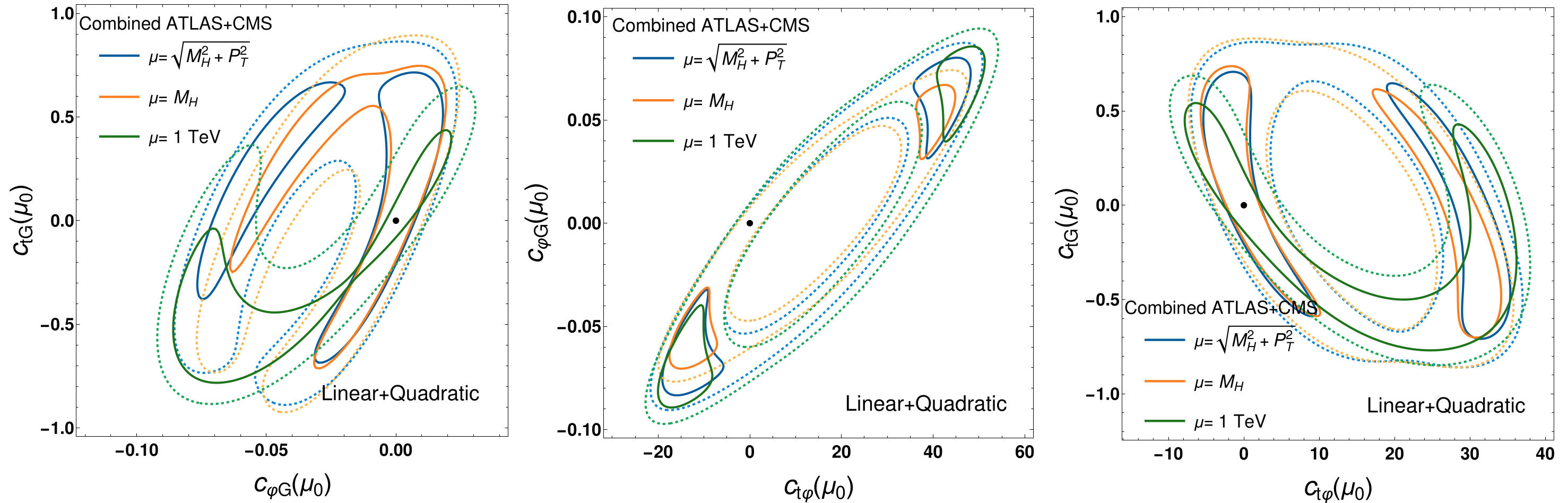}
        \caption{}
\end{subfigure}
    \caption{Two-parameters fits of the Higgs transverse momentum spectrum to combined ATLAS and CMS data in Table \ref{tab: currentdata}. In the top panel (a) only linear terms in the Wilson coefficients are considered in the computation, while in the bottom panel (b) both linear and quadratic are taken into account. Different choices of the renormalisation scale $\mu$ are considered, specifically dynamical scale scenario (blue), fixed scale (orange) and no running (green). The bounds are extracted for $\mu_0= 1$ TeV. 68\% CL and 95\% CL allowed regions are respectively indicated with solid and dashed lines.}
            \label{fig: PTH137}
\end{figure}

\paragraph{Current LHC data}
We start by showing the results of the fits for the Higgs transverse momentum spectrum. In Figure \ref{fig: PTH137} we show the allowed regions for the Wilson coefficients in a 2D parameter space, this is done by setting to zero the additional coefficient not appearing in the plot. We include the scenario where only linear terms in the Wilson coefficients are considered, as well as the one where both linear and quadratic terms are taken into account. All three combinations of coefficients are shown. 

First, we observe that differences between the different scale choices extend beyond the two-sigma level of the bounds. We notice the presence of flat directions in the SMEFT parameter space due to the observable being only sensitive to a linear combination of the Wilson coefficients \cite{Shifman:1979eb, Ellis:1975ap, 1205.1065, 1607.05330, 2109.02987, 1608.00977}. As we can observe from the plots, the scale choice impacts directly these directions, affecting the correlation between coefficients in both $\cpGG$-$\ctGG$ and $\ctpp$-$\ctGG$ planes, where the allowed directions get substantially rotated in the parameter space when running and mixing is activated.

Then, we see that when $\ctGG$ is set to zero, focusing on the $\ctpp$-$\cpGG$ plane, the impact of RG running becomes less pronounced. This is consistent with the results presented and discussed in Section \ref{sec: diffdist}, where we observed how a non-zero value of $c_{tG}$ leads to a bigger impact of the renormalisation scale choice on the SMEFT predictions, as shown in the differential distribution plots in Figure \ref{fig: Hjdistribution}.

As the inclusion of quadratic terms causes the allowed regions to deviate from an elliptic shape, we notice that for a non-zero value of the Yukawa coefficient $\ctpp$ we have a second allowed region of the parameter space away from zero. As it is well-known, this is due to $\ctpp$ appearing in the amplitude of $Hj$ and $t\bar tH$ production as a direct rescaling of the SM Yukawa coupling, leading to a degeneracy that can be seen in the 68\% CL allowed regions for $\ctpp$ when quadratic terms are included in the SMEFT analysis. In that region linear and quadratic corrections have opposite signs and cancel each other leading to predictions very close to the SM. We notice a second allowed region also for $\cpGG$ that gets excluded when considering HL-LHC projections originating from a similar cancellation of linear and quadratic contributions.

Constraints on the Wilson coefficients are obtained through individual (i.e. only one coefficient allowed at a time) and marginalised (i.e. all operators are activated) fits at 95\% C.L. shown in Table~\ref{tab: IndMargFits}, which are performed for the three choices of the renormalisation scale and including both linear and quadratic SMEFT terms.
\renewcommand{\arraystretch}{2.0}
\begin{table}[ht]
    \centering
    
    \begin{subtable}{\linewidth}
    \centering
    \resizebox{\textwidth}{!}{
    \footnotesize
    \begin{tabular}{m{1.6cm}|c|c|c}
        \hline 
        {\small Individual} & {\normalsize $\mu = \sqrt{p_T^2+M_H^2}$} & {\normalsize $\mu = M_H$} & {\normalsize $\mu= 1$ TeV (no running)} \\  
        \hline 
        \hline
        \parbox[c][0.5cm][c]{1.6cm}{\centering{\large $\ctpp$}} & [-3.58, 3.85] $\cup$ [26.14, 33.51] & [-3.37, 3.61] $\cup$ [24.69, 31.71] & [-3.86, 4.13] $\cup$ [28.18, 36.20] \\ 
        \parbox[c][0.5cm][c]{1.6cm}{\centering{\large $\cpGG$}}& [-0.074, -0.050] $\cup$ [-0.013, 0.010] & [-0.063, -0.043] $\cup$ [-0.011, 0.010] & [-0.079, -0.055] $\cup$ [-0.015, 0.011]  \\ 
        \parbox[c][0.5cm][c]{1.6cm}{\centering{\large $\ctGG$}} & [-0.29, 0.81] & [-0.26, 0.84] & [-0.21, 0.32] \\ 
        \hline 
    \end{tabular}}
    \caption{}
    \vspace{0.2cm}
\end{subtable}
    \begin{subtable}{\linewidth}
    \centering
    \resizebox{0.75\textwidth}{!}{
    \footnotesize
    \begin{tabular}{m{2.1cm}|c|c|c}
        \hline 
       {\small Marginalised} & {\normalsize $\mu = \sqrt{p_T^2+M_H^2}$} & {\normalsize $\mu = M_H$} & {\normalsize $\mu= 1$ TeV (no running)} \\  
        \hline 
        \hline
        \parbox[c][0.5cm][c]{2.1cm}{\centering{\large $\ctpp$}} & [-21.00,50.15] & [-19.56,46.98]  & [-21.17, 53.69]  \\  
        \parbox[c][0.5cm][c]{2.1cm}{\centering{\large $\cpGG$}}& [-0.095, 0.092] & [-0.085, 0.081] & [-0.10, 0.095] \\ 
        \parbox[c][0.5cm][c]{2.1cm}{\centering{\large $\ctGG$}} & [-0.68, 0.69] & [-0.70, 0.65] & [-0.77, 0.49] \\ 
        \hline 
    \end{tabular}}
    \caption{}
    \vspace{0.2cm}
\end{subtable}
    \caption{Constraints for the Wilson coefficients at 95\% C.L. (a) The individual fit results are obtained by only allowing one operator at a time, while (b) the marginalised ones are derived with all operators active. The results are shown for different choices of the renormalisation scale $\mu$. The intervals refer to the coefficients extracted at $\mu_0=1$ TeV.}
    \label{tab: IndMargFits}
\end{table}

Focusing on constraints for $\ctpp$ and $\cpGG$, both in individual and marginalised fits, a clear pattern emerges: activating the running and mixing narrows the allowed intervals by up to 20\%. The fixed-scale scenario, where the system is run down to $\mu=M_H=125$ GeV, displays the most stringent constraints among all the options. The additional interval in the individual fits, which is not compatible with zero, also shifts as the scale changes.

The constraints for the dipole coefficient $\ctGG$ exhibit a different behaviour. There is a significant variation in the allowed interval when comparing the case where running is considered to the one where it is ignored, while only minor variation is observed between the dynamical and fixed-scale scenarios. Although this holds true for both individual and marginalised fits, the latter is less affected than the former.

\paragraph{Projections for the HL-LHC}
The results for the projections for the HL-LHC are shown in Table~\ref{tab: IndMargFitsHL}. As expected, the improved precision of the measurements at the HL-LHC significantly improve all constraints. We also notice that in the individual bounds the additional interval we had for the LHC in the Yukawa coefficient is not present. This is achieved through the addition of the $M_{HH}$ distribution, as for the double Higgs production process the Yukawa operator does not lead to a pure rescaling of the SM prediction, removing the degeneracy observed when only taking into account the $p_T^H$ spectrum. Similarly the second allowed interval in the $c_{\varphi G}$ individual bounds is eliminated, due to the presence of different processes in the fit with different linear and quadratic dependences on the coefficients. 

Differences due to different scale choices for the individual bounds are similar to the ones detected also for the constraints with current data, ranging from 10-20\% for the Yukawa and contact gluon-Higgs operators and reaching a factor of 2 for $c_{tG}$. 

\begin{figure}[!]
\begin{subfigure}{\linewidth}
    \centering
     \includegraphics[width= 0.475\textwidth]{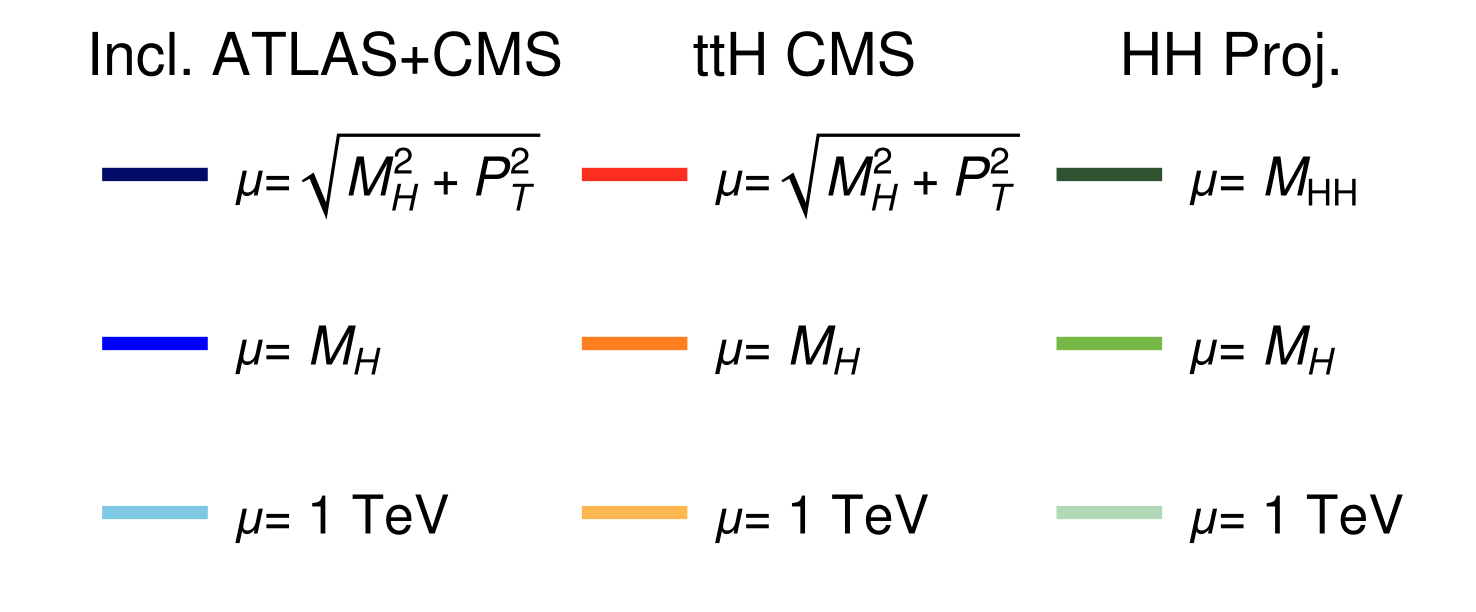}
     \vspace{0.4cm}
\end{subfigure}
\begin{subfigure}{\linewidth}
    \centering
     \includegraphics[width= \textwidth]{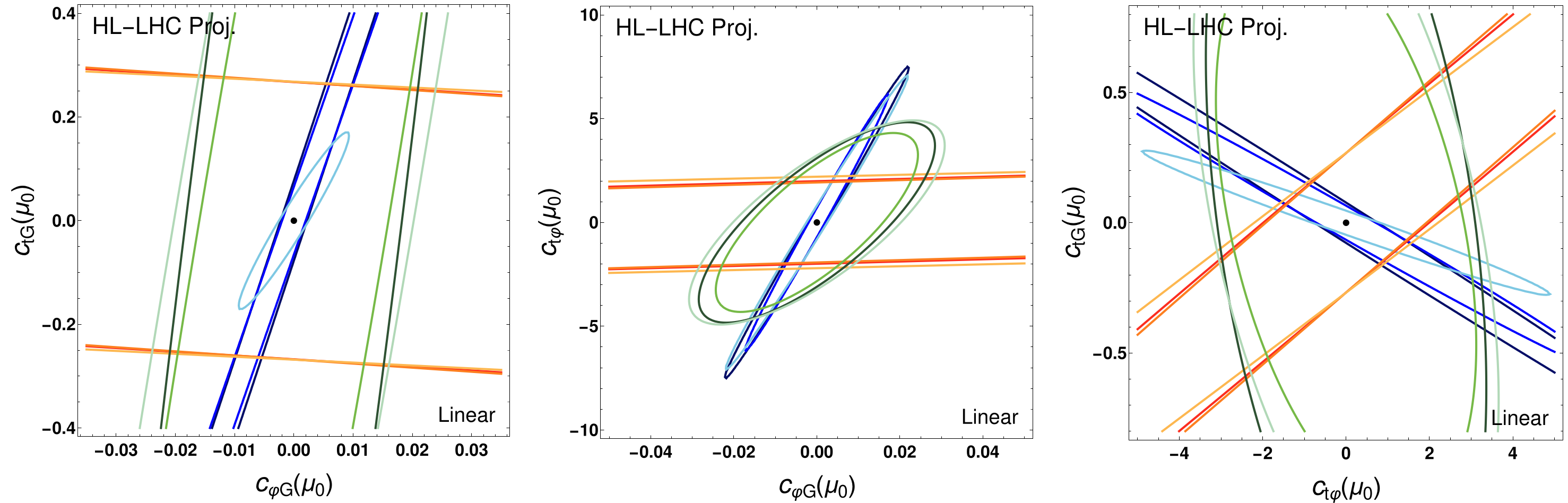}
     \caption{}
     \vspace{0.3cm}
\end{subfigure}
\begin{subfigure}{\linewidth}
    \centering
     \includegraphics[width= \textwidth]{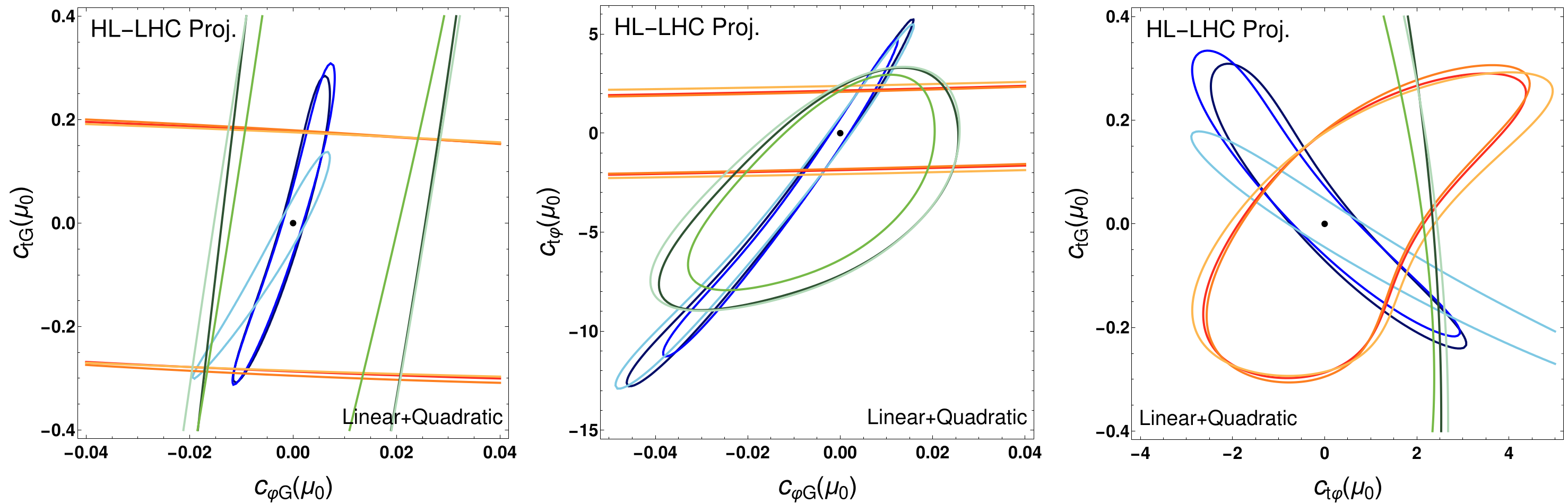}
     \caption{}
\end{subfigure}          
    \caption{HL-LHC projections for the Higgs transverse momentum spectrum with combined ATLAS and CMS uncertainties (orange), $t\bar tH$ production differential cross-section with projected uncertainties from CMS (blue) and $HH$ production (green) with uncertainties derived from the statistically dominated signal strength projection in \cite{1902.00134}.
    In panel (a) only linear terms in the Wilson coefficients are considered in the computation, while in (b) both linear and quadratic are taken into account.
    Different choices of the renormalisation scale $\mu$ are represented by different shades of the colour. The bounds are extracted for $\mu_0= 1$ TeV at 95\% C.L. }
    \label{fig: PTH3000}
\end{figure} 

Finally we show in Figure \ref{fig: PTH3000} the 2D fits for the Higgs transverse momentum spectrum both inclusive and in $t\bar tH$ production, as well as for the Higgs pair invariant mass distribution. We find that the three processes probe different directions in the SMEFT parameter space, and thus their combination helps to substantially reduce the allowed regions. In the 2D fits involving $c_{tG}$ the directions probed are significantly rotated depending on the choice of the renormalisation scale.   
\renewcommand{\arraystretch}{2.0}
\begin{table}[ht]
    \centering
    \begin{subtable}{\linewidth}
    \centering
    \footnotesize
    \begin{tabular}{m{1.6cm}|c|c|c}
        \hline 
        {\small Individual} & {\normalsize $\mu$ dynamical} & {\normalsize $\mu = M_H$} & {\normalsize $\mu= 1$ TeV (no running)} \\  
        \hline 
        \hline
        \parbox[c][0.5cm][c]{1.6cm}{\centering{\large $\ctpp$}} & [-0.55, 0.56]  & [-0.52, 0.53]  & [-0.59, 0.61]  \\ 
        \parbox[c][0.5cm][c]{1.6cm}{\centering{\large $\cpGG$}}& [-0.0018, 0.0017] & [-0.0016, 0.0015] & [-0.0020, 0.0020] \\ 
        \parbox[c][0.5cm][c]{1.6cm}{\centering{\large $\ctGG$}} & [-0.056, 0.064] & [-0.048, 0.054] & [-0.037, 0.037] \\ 
        \hline 
    \end{tabular}
    \caption{}
    \vspace{0.2cm}
\end{subtable}
    \begin{subtable}{\linewidth}
    \centering
    \footnotesize
    \begin{tabular}{m{2.1cm}|c|c|c}
        \hline 
       {\small Marginalised} & {\normalsize $\mu$ dynamical} & {\normalsize $\mu = M_H$} & {\normalsize $\mu= 1$ TeV (no running)} \\  
        \hline 
        \hline
        \parbox[c][0.5cm][c]{2.1cm}{\centering{\large $\ctpp$}} & [-2.02, 2.24] & [-1.95, 2.18]  & [-1.69, 1.59]  \\ 
        \parbox[c][0.5cm][c]{2.1cm}{\centering{\large $\cpGG$}}& [-0.012, 0.012] & [-0.012, 0.012] & [-0.010, 0.0083] \\ 
        \parbox[c][0.5cm][c]{2.1cm}{\centering{\large $\ctGG$}} & [-0.25, 0.21] & [-0.26, 0.22] & [-0.13, 0.11] \\ 
        \hline 
    \end{tabular}
    \caption{}
    \vspace{0.2cm}
\end{subtable}
    \caption{Constraints for the Wilson coefficients at 95\% C.L. with combined projected uncertainties for the HL-LHC. (a) The individual fit results are obtained by only allowing one operator at a time, while (b) the marginalised ones are derived with all operators active. The results are shown for different choices of the renormalisation scale $\mu$. The intervals refer to the coefficients extracted at $\mu_0=1$ TeV.}
    \label{tab: IndMargFitsHL}
\end{table}

\paragraph{Higgs pair production and the Higgs self-coupling}

Double Higgs production via gluon fusion is the main process to probe the Higgs trilinear coupling at hadron colliders. In the context of the SMEFT, this process receives contributions from the operators listed in Eq.~\eqref{Operators}, as well as the purely electroweak Higgs operators $O_\varphi$ and $O_{d \varphi}$, defined in Section \ref{sec: setup}. These directly impact the trilinear coupling, allowing for deviations from the SM prediction~\cite{1910.00012}. 

As purely EW operators, they do not receive corrections from QCD and therefore do not run under the QCD-induced RGEs considered in this work. However, the sensitivity of double Higgs production to the operators that run and mix under QCD may cause an indirect impact on the Higgs self-coupling operators when considering different scale scenarios. This effect would be visible in the marginalised constraints when all operators are active. In the following for simplicity we just focus on $O_\varphi$ which only enters in double Higgs production.

\begin{figure}[h]
            \centering
            \includegraphics[width= 0.9\textwidth]{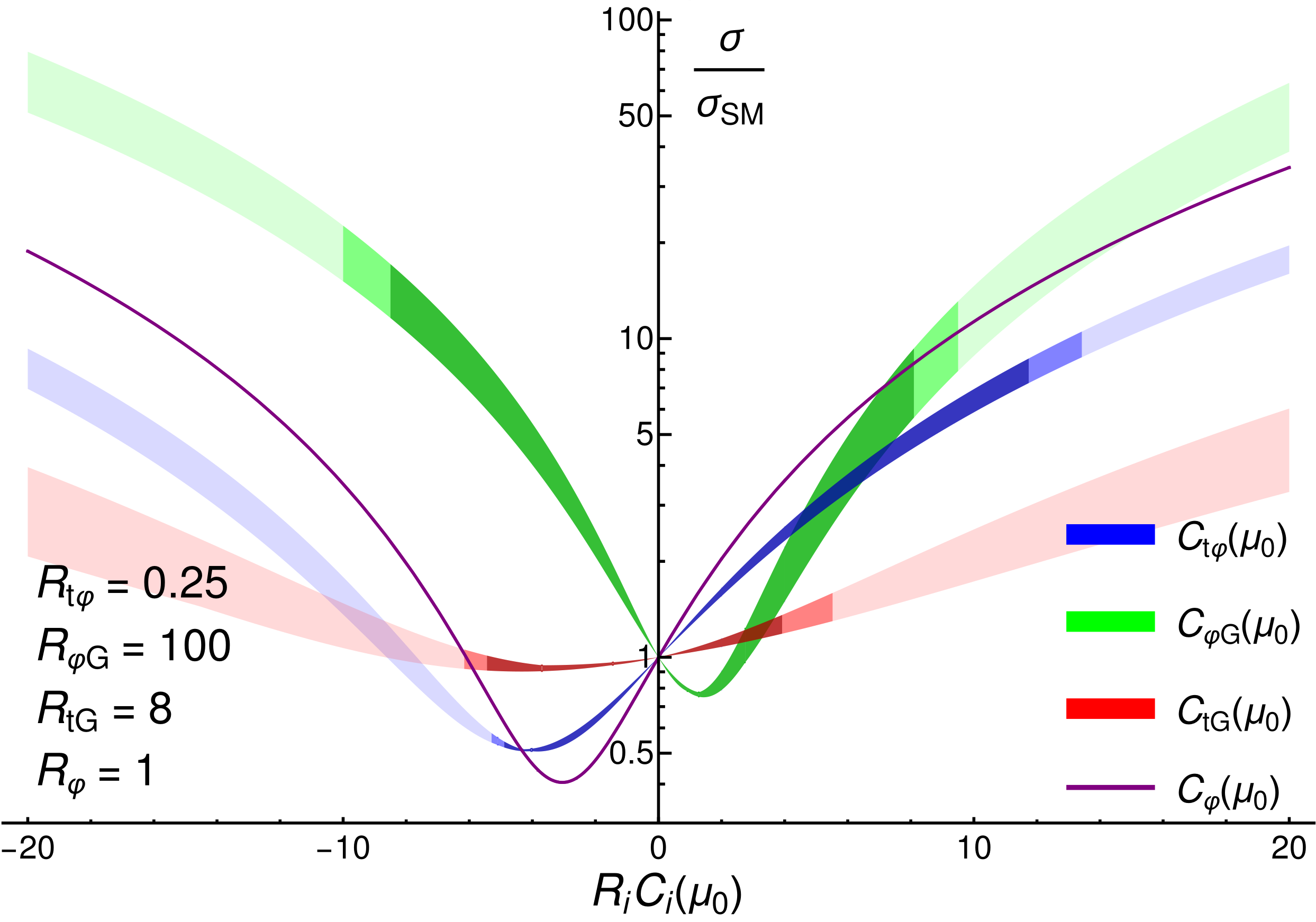}
            \caption{Double Higgs production cross-section as a function of the Wilson coefficients at the reference scale $\mu_0$ obtained considering different renormalisation scale scenarios. For plotting purposes, a different scale factor $R_i$ has been considered for each coefficient. The bands shading changes based on the marginalised limits obtained in Table \ref{tab: IndMargFits}. Darkest shading shows the allowed regions whilst the lightest shading regions is excluded by the bounds. The regions in the middle shows the variation of bounds obtained by the three running scenarios. 
   %         \FM{Bounds are not shown.}
            }
            \label{fig: butterfly}
\end{figure}

We start by showing in Figure \ref{fig: butterfly} the cross section for $HH$ production as function of the Wilson coefficients defined at the reference scale $\mu_0 = 1$ TeV. The bands indicate the variation in the total cross-section when employing different scale choices as discussed also in the rest of this work. 

\begin{figure}[h]
\begin{subfigure}{\linewidth}
    \centering
    \includegraphics[width= \textwidth]{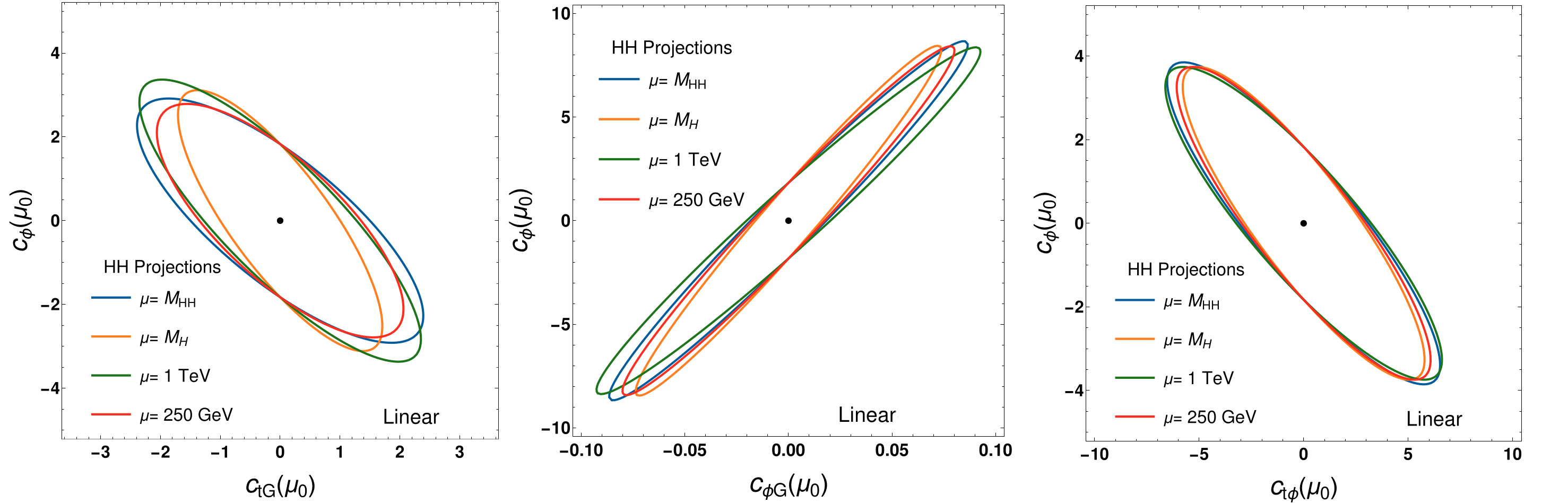}
    \caption{}
    \label{fig: cplinearFIT}
    \vspace{0.2cm}
\end{subfigure}
\begin{subfigure}{\linewidth}
    \centering
    \includegraphics[width= \textwidth]{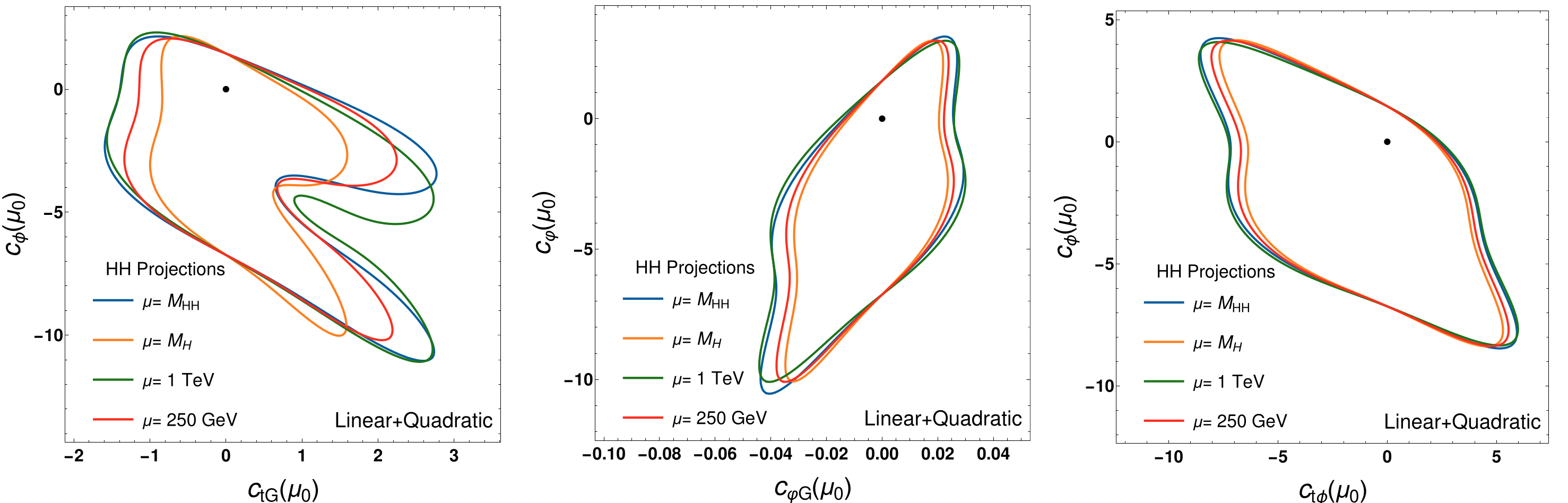}
        \caption{}
\end{subfigure}
    \caption{Two-dimensional $c_\varphi - c_i$ fit of double Higgs production at the HL-LHC to the SM spectrum. In the top panel (a) only linear terms in the Wilson coefficients are considered in the computation, while in the bottom panel (b) both linear and quadratic are taken into account. Different choices of the renormalisation scale $\mu$ are considered, specifically dynamical scale scenario (blue), fixed scale at $\mu = 125$ GeV (orange) and $\mu = 250$ GeV (red), as well as no running (green). The bounds are extracted for $\mu_0= 1$ TeV at 95\% C.L.}
            \label{fig: cpHH}
\end{figure}

Predictably, since $O_\varphi$ is insensitive to one-loop QCD corrections, activating only this operator results in no direct impact of the RGEs on the cross-section. In order to assess the impact of scale choice on constraints on $c_\varphi$ we show in Figure \ref{fig: cpHH} the allowed regions at 95\% C.L. in 2D $c_\varphi-c_i$ parameter space, by first including only linear terms and then both linear and quadratic and the $HH$ HL-LHC  projections.

When only linear terms are included, we notice an impact of the RG running/mixing on the correlation between the coefficients, where the presence of $\ctGG$ causes the biggest variation among the considered pairs of coefficients.
The inclusion of quadratic terms strongly affects the allowed regions, which extends to new values of $c_\varphi$ around a second local minimum of the $\chi^2$. Again, the biggest impact of the scale choice is due to the running and mixing of the dipole operator, with only minor variations for the other two coefficients.

Finally we show the marginalised fit for the Wilson coefficients which involves all the observables considered for the HL-LHC projections. However, this time we also include $O_\varphi$ in the computation.

\begin{table}[ht]
    \centering
    \footnotesize
    \begin{tabular}{m{2cm}|c|c|c}
        \hline 
        {\small Combined fit} & {\normalsize $\mu$ dynamical} & {\normalsize $\mu = M_H$}& {\normalsize $\mu= 1$ TeV (no running)} \\  
        \hline 
        \hline
        \parbox[c][0.5cm][c]{2cm}{\centering{\large $\ctpp$}} & [-2.03, 2.63]  & [-1.96, 2.58]  & [-1.61, 1.58]  \\ 
        \parbox[c][0.5cm][c]{2cm}{\centering{\large $\cpGG$}}& [-0.012, 0.012] & [-0.012, 0.013] & [-0.010, 0.0083] \\ 
        \parbox[c][0.5cm][c]{2cm}{\centering{\large $\ctGG$}} & [-0.26, 0.21] & [-0.27, 0.22]  & [-0.13, 0.12] \\
        \parbox[c][0.5cm][c]{2cm}{\centering{\large $c_\varphi$}} & [-6.50, 1.31] & [-6.50, 1.31] & [-6.49, 1.34] \\
        \hline 
    \end{tabular}
    \vspace{0.2cm}
    \caption{Results of marginalised fit with combined projected uncertainties for the HL-LHC including $\mathcal{O}_\varphi$. The results are shown for different choices of the renormalisation scale $\mu$. The intervals refer to the coefficients extracted with $\mu_0=1$ TeV at 95\% C.L..}
    \label{tab: IndMargFitsHL}
\end{table}

In the marginalised fit including all processes the impact of the scale choice is negligible for the bounds set on $c_{\varphi}$. We do note however that the inclusion of $\mathcal{O}_\varphi$ in the fit slightly changes the bound on the top Yukawa coefficient which shows an interplay between the single and double Higgs production processes despite the fact that double Higgs production has a significantly smaller cross-section.  

\section{Conclusions}
\label{sec: conclusions}
In this work we studied the impact of running and mixing of Wilson coefficients in the SMEFT on Higgs observables. We focused on three SMEFT operators relevant for Higgs production, a set closed under the QCD-induced renormalisation, and implemented the renormalisation group flow through a reweighting technique based on the cross-section results obtained with \texttt{MadGraph5\_aMC@NLO} using the model \texttt{SMEFT@NLO}.

We analysed in detail the Higgs transverse momentum and double Higgs invariant mass spectra which required SMEFT analysis of the loop-induced $Hj$ and $HH$ production via gluon fusion, as well as $t\bar tH$ production. We studied the impact of scale choice on differential observables considering both fixed and dynamical scale scenarios, compared with the case where running is ignored. These effects were found to be of $\mathcal{O}(10-40\%)$ when the chromomagnetic dipole operator present at high scales, and up to $\mathcal{O}(50\%)$ when it was set to a non-zero value at the reference scale due to a strong mixing with the gluon-Higgs operator, and combined with the $\mu$-dependent cross-section terms arising upon renormalising the corresponding amplitudes. 

In the particular case of a non-zero chromomagnetic coefficient we have decomposed the SMEFT cross-section into the three components associated to the three operators for the linear contributions and to six associated contributions in the squared ones, revealing interesting patterns and cancellations between different constributions for the loop induced processes. Mixing generated contributions due to the induced $c_{\varphi G}$ coefficient dominate in the cross-section at lower energies. 

Benefiting from the existence of experimental data for the $p_T^H$ spectrum at the LHC Run II, we studied the impact of RG on constraints obtained for the Wilson coefficients through a toy fit to data. We found the RG effects to exceed the two-sigma level of the bounds with a significant impact on the correlations between the Wilson coefficients, resulting in flat directions being rotated in the SMEFT parameter space. 

We then employed projections for the HL-LHC using projected uncertainties available in the literature, and separately including $t\bar tH$ differential cross-sections, as well as $HH$ production finding again a significant impact of RG evolution on the correlations between coefficients. 

In both cases, we performed both individual and marginalised fits to extract bounds for the coefficients finding significant variation of the allowed intervals in particular for the operators which generate new operators through mixing. In general, we have demonstrated that the impact of scale choice on the allowed intervals is not negligible and running and mixing should be taken into account. In the case of Higgs pair production we also explored the impact of RG effects in a marginalised fit involving both operators that run under QCD and the operators modifying the Higgs self-coupling.  

Our study, performed for a small set of observables and operators, is not an attempt to set realistic constraints on these Wilson coefficients. In particular we have selected a small set of differential observables, for which a dynanical scale choice is more natural and for which predictions vary most once running and mixing is taken into consideration. In fact, setting more realistic constraints can only be achieved in global fits with the inclusion of additional processes, inclusive and differential, and operators. These global interpretations demand also a proper treatment of various theoretical and experimental uncertainties. Our motivation was to determine how, in a sufficiently realistic scenario, the scale dependence would impact such constraints with the final goal of encouraging the proper inclusion of RG effects in global fits analyses. As more precise and more differential experimental data are expected in the future, our results show that including these RG effects in global fits will be imperative to consistently and reliably extract constraints on new physics.

\bibliographystyle{JHEP.bst}%Used BibTeX style is unsrt
\bibliography{paper.bib}
\end{document}

%% file: Fdiagrams.tex
\begin{figure}[h]
    \centering
\begin{subfigure}{\textwidth}
\centering
\begin{centering}
\scalebox{1}{
\begin{tikzpicture}
\begin{feynman}
\vertex (i1) {$g$};
\vertex[below=1.2cm of i1] (i2) {$g$};
\vertex[right=1cm of i1, dot, minimum size=2.2mm, red] (l1) {};
\vertex[below=1.2cm of l1, dot, minimum size=2.2mm, red] (l3) {};
\vertex[below right=0.6cm and 1cm of l1,  dot, minimum size=2.2mm, blue] (l2) {} ;
\vertex[right=0.65cm of l2, empty dot] (v2) ;
\vertex[right=2.6cm of l1] (f1) {$H$};
\vertex[right=2.6cm of l3] (f2) {$H$};
\diagram* {{[edges={fermion, arrow size=1pt}]
(l1) -- (l2) -- (l3) -- (l1),},
(i1) -- [gluon] (l1),
(i2) -- [gluon] (l3),
(l2) -- [scalar] (v2),
(v2) -- [scalar] (f1),
(v2) -- [scalar] (f2) };
\end{feynman}
\end{tikzpicture}}
\scalebox{1}{
\begin{tikzpicture}
\begin{feynman}
\vertex (i1) {$g$};
\vertex[below=1.2cm of i1] (i2) {$g$};
\vertex[right=1cm of i1] (l1);
\vertex[right=1.2cm of l1] (l2);
\vertex[below right = 1.1cm and 0.6cm of l1, dot, minimum size=2.2mm, red] (l3) {};
\vertex[right=0.6cm of l2] (f2) {$H$};
\vertex[below=1.2cm of f2] (f1) {$H$};
\diagram* {{[edges={fermion, arrow size=1pt}]
(l1) -- (l2) -- (l3) -- (l1),},
(i1) -- [gluon] (l1),
(i2) -- [gluon] (l3),
(l3) -- [scalar] (f1),
(l2) -- [scalar] (f2) };
\end{feynman}
\end{tikzpicture}}
%%%%% ctg topology, gg to blob to H propagator, ggtt coupling.
\scalebox{1}{
\begin{tikzpicture}
\begin{feynman}
\vertex (i1) {$g$};
\vertex[below=1.2cm of i1] (i2) {$g$};
\vertex[right=1.1cm of i1] (l1);
\vertex[dot, minimum size=2.2mm, red] (ph1) [right=1.1cm of i1] {};
\vertex[right=1.1cm of i2] (l2);
\vertex[below right=0.5 cm and 0.866cm of l1] (v1);
\vertex[above right=0.1cm and 2.6cm of i1] (f1) {$H$};
\vertex[right=2.6cm of i2] (f2) {$H$};
\diagram* {{[edges={fermion, arrow size=1pt}]
(l1) -- [out=0, in=0] (l2) -- [out=180, in=180] (l1),},
(i1) -- [gluon] (l1),
(i2) -- [gluon] (l2),
(l1) -- [scalar] (v1),
(v1) -- [scalar] (f1),
(v1) -- [scalar] (f2) };
\end{feynman}
\end{tikzpicture}}
\scalebox{1}{
\begin{tikzpicture}
\begin{feynman}
\vertex (i1) {$g$};
\vertex[below=1.2cm of i1] (i2) {$g$};
\vertex[right=1 of i1] (l1);
\vertex[below=1.2cm of l1] (l3);
\vertex[below right=0.6cm and 1cm of l1, dot, minimum size=2.2mm, blue] (l2) {};
\vertex[right=2cm of l1] (f1) {$H$};
\vertex[right=2cm of l3] (f2) {$H$};
\diagram* {{[edges={fermion, arrow size=1pt}]
(l1) -- (l2) -- (l3) -- (l1),},
(i1) -- [gluon] (l1),
(i2) -- [gluon] (l3),
(l2) -- [scalar] (f1),
(l2) -- [scalar] (f2) };
\end{feynman}
\end{tikzpicture}}
\scalebox{1}{
\begin{tikzpicture}
\begin{feynman}
\vertex (i1) {$g$};
\vertex[below=1.2cm of i1] (i2) {$g$};
\vertex[right=1.2cm of i1, dot, minimum size=2.2mm, red] (l1) {};
\vertex[right=1.2cm of l1, dot, minimum size=2.2mm, blue] (l2) {};
\vertex[below=1.2cm of l2, dot, minimum size=2.2mm, blue] (l3) {};
\vertex[below=1.2cm of l1, dot, minimum size=2.2mm, red] (l4) {};
\vertex[right=1.2cm of l3] (f1) {$H$};
\vertex[right=1.2cm of l2] (f2) {$H$};
\diagram* {{[edges={fermion, arrow size=1pt}]
(l1) -- (l2) -- (l3) -- (l4) -- (l1),},
(i1) -- [gluon] (l1),
(i2) -- [gluon] (l4),
(l3) -- [scalar] (f1),
(l2) -- [scalar] (f2) };
\end{feynman}
\end{tikzpicture}}
\scalebox{1}{
\begin{tikzpicture}
\begin{feynman}
\vertex (i1) {$g$};
\vertex[below=1.4cm of i1] (i2) {$g$};
\vertex[below right=0.7cm and 1cm of i1, dot, minimum size=2.2mm, mygreen] (v1) {};
\vertex[right = 1cm of v1] (v2);
\vertex[right=2.6cm of i1] (f1) {$H$};
\vertex[right=2.6cm of i2] (f2) {$H$};
\diagram* {
(i1) -- [gluon] (v1),
(i2) -- [gluon] (v1),
(v1) -- [scalar] (v2),
(v2) -- [scalar] (f1),
(v2) -- [scalar] (f2) };
\end{feynman}
\end{tikzpicture}}
\scalebox{1}{
\begin{tikzpicture}
\begin{feynman}
\vertex (i1) {$g$};
\vertex[below=1.4cm of i1] (i2) {$g$};
\vertex[below right=0.7cm and 1cm of i1, dot, minimum size=2.2mm, mygreen] (v1) {};
\vertex[right=2cm of i1] (f1) {$H$};
\vertex[right=2cm of i2] (f2) {$H$};
\diagram* {
(i1) -- [gluon] (v1),
(i2) -- [gluon] (v1),
(v1) -- [scalar] (f1),
(v1) -- [scalar] (f2) };
\end{feynman}
\end{tikzpicture}}
\end{centering}
\caption{}
\end{subfigure}
\begin{subfigure}{\textwidth}
\centering
    \begin{centering}
\scalebox{1}{
\begin{tikzpicture}
\begin{feynman}
\vertex (i1) {$g$};
\vertex[below=1.2cm of i1] (i2) {$g$};
\vertex[right=1cm of i1] (l1);
\vertex[right=1.2cm of l1] (l2);
\vertex[below right = 1.1cm and 0.6cm of l1, dot, minimum size=2.2mm, red] (l3) {};
\vertex[right=0.7cm of l2] (f2) {$g$};
\vertex[below=1.2cm of f2] (f1) {$H$};
\diagram* {{[edges={fermion, arrow size=1pt}]
(l1) -- (l2) -- (l3) -- (l1),},
(i1) -- [gluon] (l1),
(i2) -- [gluon] (l3),
(l3) -- [scalar] (f1),
(l2) -- [gluon] (f2) };
\end{feynman}
\end{tikzpicture}}
\scalebox{1}{
\begin{tikzpicture}
\begin{feynman}
\vertex (i1) {$g$};
\vertex[below=1.2cm of i1] (i2) {$g$};
\vertex[right=1cm of i1] (l1);
\vertex[below=1.2cm of l1] (l3);
\vertex[below right=0.6cm and 1cm of l1, dot, minimum size=2.2mm, red] (l2) {};
\vertex[right=2cm of l1] (f1) {$g$};
\vertex[right=2cm of l3] (f2) {$H$};
\diagram* {{[edges={fermion, arrow size=1pt}]
(l1) -- (l2) -- (l3) -- (l1),},
(i1) -- [gluon] (l1),
(i2) -- [gluon] (l3),
(l2) -- [gluon] (f1),
(l2) -- [scalar] (f2) };
\end{feynman}
\end{tikzpicture}}

\scalebox{1}{
\begin{tikzpicture}
\begin{feynman}
\vertex (i1) {$g$};
\vertex[below=1.2cm of i1] (i2) {$g$};
\vertex[right=1.2cm of i1, dot, minimum size=2.2mm, red] (l1) {};
\vertex[right=1.2cm of l1, dot, minimum size=2.2mm, blue] (l2) {};
\vertex[below=1.2cm of l2, dot, minimum size=2.2mm, red] (l3) {};
\vertex[below=1.2cm of l1, dot, minimum size=2.2mm, red] (l4) {};
\vertex[right=1.2cm of l3] (f1) {$g$};
\vertex[right=1.2cm of l2] (f2) {$H$};
\diagram* {{[edges={fermion, arrow size=1pt}]
(l1) -- (l2) -- (l3) -- (l4) -- (l1),},
(i1) -- [gluon] (l1),
(i2) -- [gluon] (l4),
(l3) -- [gluon] (f1),
(l2) -- [scalar] (f2) };
\end{feynman}
\end{tikzpicture}}
\scalebox{1}{
\begin{tikzpicture}
\begin{feynman}
\vertex (i1) {$g$};
\vertex[below=1.4cm of i1] (i2) {$g$};
\vertex[below right=0.7cm and 1cm of i1, dot, minimum size=2.2mm, mygreen] (v1) ;
\vertex[right = 1cm of v1, dot, minimum size=2.2mm, mygreen] (v2) {};
\vertex[right=2.8cm of i1] (f1) {$g$};
\vertex[right=2.8cm of i2] (f2) {$H$};
\diagram* {
(i1) -- [gluon] (v1),
(i2) -- [gluon] (v1),
(v1) -- [gluon] (v2),
(v2) -- [gluon] (f1),
(v2) -- [scalar] (f2) };
\end{feynman}
\end{tikzpicture}}
\scalebox{1}{
\begin{tikzpicture}
\begin{feynman}
\vertex (i1) {$g$};
\vertex[below=1.4cm of i1] (i2) {$g$};
\vertex[below right=0.7cm and 1cm of i1, dot, minimum size=2.2mm, mygreen] (v1) {};
\vertex[right=2cm of i1] (f1) {$g$};
\vertex[right=2cm of i2] (f2) {$H$};
\diagram* {
(i1) -- [gluon] (v1),
(i2) -- [gluon] (v1),
(v1) -- [gluon] (f1),
(v1) -- [scalar] (f2) };
\end{feynman}
\end{tikzpicture}}
\end{centering}
\caption{}
\end{subfigure}

\begin{subfigure}{\textwidth}
    \centering
    \begin{centering}
        \scalebox{1}{
\begin{tikzpicture}
\begin{feynman}
\vertex (i1) {$g$};
\vertex[below=1.2cm of i1] (i2) {$g$};
\vertex[right=1.2cm of i1, dot,minimum size=2.2mm, red] (l1) {};
\vertex[below=1.2cm of l1, dot,minimum size=2.2mm, red] (l3) {};
\vertex[below=0.6cm of l1,  dot,minimum size=2.2mm, blue] (l2) {} ;
\vertex[right=1.3cm of l1] (f1) {$t$};
\vertex[right=1.3cm of l2] (f2) {$H$};
\vertex[right=1.3cm of l3] (f3) {$\Bar{t}$};
\diagram* {
(i1) -- [gluon] (l1) -- [fermion, arrow size=1pt] (f1),
(i2) -- [gluon] (l3) -- [anti fermion, arrow size=1pt] (f3),
(l2) -- [scalar] (f2),
(l1) -- (l2) -- (l3)
};
\end{feynman}
\end{tikzpicture}}
\scalebox{1}{
\begin{tikzpicture}
\begin{feynman}
\vertex (i1) {$g$};
\vertex[below=1.2cm of i1] (i2) {$g$};
\vertex[below right=0.6cm and 1.1cm of i1, dot,minimum size=2.2mm, red] (v1) {};
\vertex[right=2.2cm of i1] (f1) {$t$};
\vertex[below right=0.6cm and 2.4cm of i1] (f2) {$H$};
\vertex[right=2.2cm of i2] (f3) {$\Bar{t}$};
\diagram* {
(i1) -- [gluon] (v1),
(i2) -- [gluon] (v1) ,
(v1) -- [fermion, arrow size=1pt] (f1),
(v1) -- [scalar] (f2),
(v1) -- [anti fermion, arrow size=1pt] (f3)
};
\end{feynman}
\end{tikzpicture}}
\scalebox{1}{
\begin{tikzpicture}
\begin{feynman}
\vertex (i1) {$g$};
\vertex[below=1.3cm of i1] (i2) {$g$};
\vertex[right=1.2cm of i1, dot,minimum size=2.2mm, blue] (l1) ;
\vertex[below=1.2cm of l1, dot,minimum size=2.2mm,  mygreen] (l3) {} ;
\vertex[below=0.6cm of l1,  dot,minimum size=2.2mm, mygreen] (l2) ;
\vertex[right=1.3cm of l1] (f1) {$t$};
\vertex[right=1.3cm of l2] (f2) {$\Bar{t}$};
\vertex[right=1.5cm of l3] (f3) {$H$};
\diagram* {
(i1) -- [gluon] (l1) -- [fermion, arrow size=1pt] (f1),
(i2) -- [gluon] (l3) -- [scalar] (f3),
(l2) -- [anti fermion, arrow size=1pt] (f2),
(l1) -- (l2) -- [gluon] (l3)
};
\end{feynman}
\end{tikzpicture}}
\scalebox{1}{
\begin{tikzpicture}
\begin{feynman}
\vertex (i1) {$g$};
\vertex[below=1.2cm of i1] (i2) {$g$};
\vertex[below right=0.6cm and 1.1cm of i1, dot,minimum size=2.2mm, mygreen] (v1) {};
\vertex[below right= 0.2cm and 1.8cm of i1] (v2);
\vertex[above right =0.2cm and 0.5cm of v2] (f1) {$t$};
\vertex[below right =0.2cm and 0.5cm of v2] (f2) {$\Bar{t}$};
\vertex[right=2.2cm of i2] (f3) {$H$};
\diagram* {
(i1) -- [gluon] (v1),
(i2) -- [gluon] (v1) ,
(v1) -- [gluon] (v2),
(v1) -- [scalar] (f3),
(v2) -- [fermion, arrow size=1pt] (f1),
(v2) -- [anti fermion, arrow size=1pt] (f2)
};
\end{feynman}
\end{tikzpicture}}

\scalebox{1}{
\begin{tikzpicture}
\begin{feynman}
\vertex (i1) {$q$};
\vertex[below=1.4cm of i1] (i2) {$\Bar{q}$};
\vertex[below right=0.7cm and 0.9cm of i1, dot,minimum size=2.2mm, mygreen] (v1);
\vertex[right=0.8cm of v1, dot,minimum size=2.2mm, mygreen] (v3) {};
\vertex[above right= 0.35cm and 0.6cm of v3, dot,minimum size=2.2mm, red] (v2);
\vertex[above right =0.3cm and 0.7cm of v2] (f1) {$t$};
\vertex[below right =0.2cm and 0.7cm of v2] (f2) {$\Bar{t}$};
\vertex[right=3cm of i2] (f3) {$H$};
\diagram* {
(i1) -- [fermion, arrow size=1pt] (v1),
(i2) -- [anti fermion, arrow size=1pt] (v1) ,
(v1) -- [gluon] (v3),
(v3) -- [scalar] (f3),
(v3) -- [gluon] (v2),
(v2) -- [fermion, arrow size=1pt] (f1),
(v2) -- [anti fermion, arrow size=1pt] (f2)
};
\end{feynman}
\end{tikzpicture}}
\scalebox{1}{
\begin{tikzpicture}
\begin{feynman}
\vertex (i1) {$q$};
\vertex[below=1.4cm of i1] (i2) {$\Bar{q}$};
\vertex[below right=0.7cm and 0.9cm of i1, dot,minimum size=2.2mm, mygreen] (v1);
\vertex[right=0.8cm of v1, dot,minimum size=2.2mm, red] (v3) {};
\vertex[above right= 0.35cm and 0.6cm of v3, dot,minimum size=2.2mm, blue] (v2) {};
\vertex[above right =0.45cm and 0.7cm of v2] (f1) {$t$};
\vertex[below right =0.35cm and 0.7cm of v2] (f2) {$H$};
\vertex[right=3cm of i2] (f3) {$\Bar{t}$};
\diagram* {
(i1) -- [fermion, arrow size=1pt] (v1),
(i2) -- [anti fermion, arrow size=1pt] (v1) ,
(v1) -- [gluon] (v3),
(v3) -- [anti fermion, arrow size=1pt] (f3),
(v3) -- [fermion, arrow size=1pt] (v2),
(v2) -- [fermion, arrow size=1pt] (f1),
(v2) -- [scalar] (f2)
};
\end{feynman}
\end{tikzpicture}}
\scalebox{1}{
\begin{tikzpicture}
\begin{feynman}
\vertex (i1) {$q$};
\vertex[below=1.4cm of i1] (i2) {$\Bar{q}$};
\vertex[below right=0.7cm and 0.9cm of i1, dot,minimum size=2.2mm, blue] (v2);
\vertex[right=0.8cm of v2, dot,minimum size=2.2mm, red] (v1) {};
\vertex[right=2.8cm of i1] (f1) {$t$};
\vertex[below right=0.7cm and 3.0cm of i1] (f2) {$H$};
\vertex[right=2.8cm of i2] (f3) {$\Bar{t}$};
\diagram* {
(i1) -- [fermion, arrow size=1pt] (v2),
(i2) -- [anti fermion, arrow size=1pt] (v2) ,
(v2) -- [gluon] (v1) , 
(v1) -- [fermion, arrow size=1pt] (f1),
(v1) -- [scalar] (f2),
(v1) -- [anti fermion, arrow size=1pt] (f3)
};
\end{feynman}
\end{tikzpicture}}
    \end{centering}
    \caption{}
\end{subfigure}
\caption{Example Feynman diagrams for $HH$ (a), $Hj$ (b) and $t\bar{t}H$ (c) production at the LHC with insertions of SMEFT operators. $O_{t\varphi}$ insertion is indicated in blue, $O_{\varphi G}$ in green and $O_{tG}$ in red. For diagrams with multiple dots,  one insertion at a time is considered in the computation.}
\label{fig: FeynmanDiagrams}
\end{figure}